%% file: SUS-11-016_temp.tex
\pdfoutput=1

\documentclass[11pt,twoside,a4paper,cmspaper,final,collab]{cms-tdr}
\def\svnVersion{218722}\def\cmsCernNo{2012-351}\def\cmsCernDate{\today}\def\cmsMessage{Published in Physical Review D as \href{http://dx.doi.org/10.1103/PhysRevD.88.052017}{\doi{10.1103/PhysRevD.88.052017.}}}

\begin{document}\cmsNoteHeader{SUS-11-016}

\hyphenation{had-ron-i-za-tion}
\hyphenation{cal-or-i-me-ter}
\hyphenation{de-vices}

\RCS$Revision: 205347 $
\RCS$HeadURL: svn+ssh://svn.cern.ch/reps/tdr2/papers/SUS-11-016/trunk/SUS-11-016.tex $
\RCS$Id: SUS-11-016.tex 205347 2013-09-02 12:45:44Z alverson $
\newcommand{\HsT}{\ensuremath{\not\!\!\mathrm{H}_{\mathrm{T}}}\xspace}
\newcommand{\HsTjets}{\ensuremath{\not\!\!\mathrm{H}_{\mathrm{T}}}+jets\xspace}
\newcommand{\MTtwo}{\ensuremath{{M}_{\mathrm{T2}}}\xspace}
\newcommand{\MTtwob}{\ensuremath{{M}_{\mathrm{T2}}\cPqb}\xspace}
\newcommand{\AlphaT}{\ensuremath{\alpha_{\mathrm{T}}}\xspace}
\newcommand{\SigNLL}{\ensuremath{\sigma^{\mathrm{NLO+NLL}}}\xspace}
\newcommand{\SigNLO}{\ensuremath{\sigma^{\mathrm{NLO}}}\xspace}
\providecommand{\fb} {\ensuremath{\,\text{fb}}\xspace}
\newcommand\sq\PSQ
\newcommand\gl\PSg
\newcommand\chioz{\ensuremath{\widetilde{\chi}_{\tiny \mathrm{LSP}}}\xspace} 
\newcommand\chioc{\ensuremath{\widetilde{\chi}^\pm_1}\xspace}
\renewcommand\chipm{\ensuremath{\widetilde{\chi}^\pm_1}\xspace}
\newcommand\chitz{\PSGczDt}
\newcommand\chithreez{\ensuremath{\widetilde{\chi}^0_3}\xspace}
\providecommand{\qqbar}{\ensuremath{\cmsSymbolFace{q}\overline{\cmsSymbolFace{q}}}\xspace}
\newcommand{\accXeff}{\ensuremath{\mathcal{A}\times \epsilon}\xspace}
\newcommand{\sigmaXBF}{\ensuremath{\sigma\times\mathcal{B}}\xspace}
\newcommand{\sigmaXBFUL}{\ensuremath{\left[{\sigma\times\mathcal{B}}\right]_\mathrm{UL}}\xspace}

\newcommand{\emu}{\Pe/\Pgm\xspace}

\newcommand{\ymet}{\ensuremath{\mathrm{Y}_{\mathrm{MET}}}\xspace}

\newcommand\tablespace{\rule[-1ex]{0cm}{3.5ex}}

\title{Interpretation of searches for supersymmetry with simplified models}

\date{\today}

\abstract{
The results of searches for supersymmetry by the CMS experiment
are interpreted in the framework of  simplified models.
The results are based on data corresponding
to an integrated luminosity of 4.73 to 4.98\fbinv.
The data were collected at the LHC
in proton--proton collisions at a center-of-mass energy
of 7\TeV.
This paper describes the method of
interpretation and provides
upper limits on the product of the production cross section and branching fraction
as a function of new particle masses for a number of simplified models.
These limits and the
corresponding experimental acceptance calculations can
be used to constrain other theoretical models and to compare different
supersymmetry-inspired analyses.
}

\hypersetup{%
pdfauthor={CMS Collaboration},%
pdftitle={Interpretation of searches for supersymmetry with simplified models},%
pdfsubject={CMS},%
pdfkeywords={CMS, physics, supersymmetry, simplified models}}

\maketitle %maketitle comes after all the front information has been supplied

\section{Introduction}
\label{sec:introduction}

The results of searches for supersymmetry (SUSY)~\cite{Wess:1974tw} at particle colliders
are often used to test the validity of a few, specific, theoretical models.
These models predict a large number of experimental observables at hadron colliders
as a function of a few theoretical parameters.
Most of the SUSY analyses performed by the Compact Muon Solenoid (CMS) experiment
present their results as an exclusion of a range of parameters for
the constrained minimal
supersymmetric standard model (CMSSM)~\cite{ref:MSUGRA,PhysRevLett.69.725,cmssm}.
However, the results of the SUSY analyses
can be used to test a wide range of alternative models, since many SUSY and non-SUSY
models predict a similar phenomenology.
These similarities inspired
the formulation
of the simplified model framework for presenting experimental results~\cite{ArkaniHamed:2007fw,Alwall:2008ag,Alwall:2008va,Alves:2011sw,Alves:2011wf}.
Specific applications of these ideas have appeared in Refs.~\cite{Papucci:20113g,Mahbubani:2012qq}.

A simplified model is defined by a set of hypothetical
particles and a sequence of their production and decay.
For each simplified model, values
for the product of the experimental acceptance and efficiency ($\accXeff$)
are calculated to translate a number of signal events
into a signal cross section.
From this information,
a 95\% confidence level upper limit (UL) on
the product of the cross section and branching fraction
(\sigmaXBFUL) is derived
as a function of particle masses.
The simplified model framework can quantify the dependence of an experimental limit on the particle spectrum or a particular sequence of particle production and decay in a manner that is more general than the CMSSM.
Furthermore, the values of \sigmaXBFUL can be compared with
theoretical predictions from a SUSY or non-SUSY model
to determine whether the theory is compatible with data.

This paper collects and describes simplified model interpretations of
a large number of SUSY-inspired analyses performed on data collected by the CMS collaboration in 2011~\cite{SUS-11-022,SUS-12-011,SUS-12-002,SUS-12-003,
SUS-11-024,
SUS-12-010,SUS-11-028,
SUS-11-010,SUS-11-020,SUS-11-011,SUS-11-018,SUS-11-013,SUS-12-006,SUS-11-021,SUS-12-001}.
The simplified model framework was also applied by CMS to a limited number of analyses in 2010~\cite{SUS-10-005}.
The ATLAS collaboration has published similar
interpretations~\cite{ATLAS-PAPERS-SUSY-2012-192,ATLAS-PAPERS-SUSY-2012-194,ATLAS-PAPERS-SUSY-2012-216,ATLAS-PAPERS-SUSY-2012-217,ATLAS-PAPERS-SUSY-2012-052,ATLAS-PAPERS-SUSY-2012-141,ATLAS-PAPERS-2012-204}.

The paper is organized as follows:
Section~\ref{sec:analyses} provides a brief description
of the CMS analyses considered here; Section~\ref{sec:simplifiedmodels} describes simplified models;
Section~\ref{sec:procedure} demonstrates the calculation of the product of the experimental acceptance and efficiency and
the upper limits on cross sections;
Section~\ref{sec:comparisons} contains comparisons
of the results for different simplified models and analyses; Section~\ref{sec:conclusions}
contains a summary.

\section{The CMS detector and analyses}
\label{sec:analyses}

The CMS detector consists of a silicon tracker, an electromagnetic calorimeter, and a hadronic calorimeter,
all located within the field volume of a central solenoid magnet, and a muon-detection system located outside the magnet~\cite{cms}.
Information from these components is combined to define
objects such as
electrons, muons, photons, jets, jets identified as \cPqb\ jets (\cPqb-tagged jets), and missing transverse energy (\ETslash).
The exact definition of these objects depends on the specific analysis, and can be found in the analysis
references.
The data were collected by the CMS experiment at the Large Hadron Collider in
proton-proton collisions
at a center-of-mass energy of 7\TeV.
Unless stated otherwise,
the data corresponds to an integrated luminosity of $4.98\pm 0.11\fbinv$~\cite{CMS-PAS-SMP-12-008}.

The descriptions of the analyses are categorized by
the main features of the event selection.
Detailed descriptions of these analyses can be found in the references~\cite{SUS-11-022,SUS-12-011,SUS-12-002,SUS-12-003,
SUS-11-024,
SUS-12-010,SUS-11-028,
SUS-11-010,SUS-11-020,SUS-11-011,SUS-11-018,SUS-11-013,SUS-12-006,SUS-11-021,SUS-12-001}.
The target of these analyses is a signal of the production of new, heavy
particles that decay into standard model particles and
stable, neutral particles that escape detection.
The stable, neutral particles can produce a signature of large \ETslash.
The standard model also produces \ETslash in top quark, weak gauge boson,
and heavy flavor production.   Fluctuations in energy deposition in the detector can
also produce significant \ETslash in quantum chromodynamics processes.

\begin{description}

\item[All-Hadronic]  Events contain two or more high transverse momentum (\pt) jets and significant \ETslash.
Events with isolated leptons are rejected to reduce
backgrounds from \ttbar, \PW, and \Z boson production.
A selection on kinematic discriminants is applied to reduce backgrounds containing \ETslash.
The names of the discriminants label the analyses:
\AlphaT~\cite{SUS-11-022}, \HsTjets~\cite{SUS-12-011} and \MTtwo~\cite{SUS-12-002}.
The \AlphaT~\cite{Randall:2008rw} and \MTtwo~\cite{Barr:2003rg,Lester:1999tx} variables are both motivated by the kinematics of new-particle
pair production and decay into two visible systems of jets and a pair of invisible particles.
The \HsTjets analysis, instead, uses a selection on the negative vector sum (\HsT) and
the scalar sum (\HT) of the transverse momentum of each jet.
The \MTtwo analysis uses data corresponding to an integrated luminosity of $4.73\pm 0.10\fbinv$

The \AlphaT analysis mentioned above also categorizes
events with one, two, or at least three jets that satisfy a \cPqb-tagging requirement.
The \MTtwob analysis modifies the \MTtwo selection mentioned above
and requires at least
one \cPqb-tagged jet.
The  $\ETslash+\cPqb$ analysis~\cite{SUS-12-003} follows a similar
strategy as the \HsTjets analysis, but selects events with one, two, or at least three \cPqb-tagged jets.

\item[Single Lepton + Jets] Events are selected with
one high-\pt, isolated lepton (electron or muon), jets, and significant \ETslash.
Three analyses are considered in this paper~\cite{SUS-12-010}.
The lepton spectrum (\emu LS) and lepton projection (\emu LP) methods exploit the
expected correlation between the lepton \pt and \ETslash from
\PW\ boson decays to separate a potential signal from the main backgrounds
of \ttbar and \PW+jets production.
The artificial neutral network (ANN) method applies an ANN that is based on
event properties (jet multiplicity, \HT, transverse mass, and the
azimuthal angular separation between the two highest-\pt jets) to separate
backgrounds from the expectations of a CMSSM benchmark model.

Two other analyses~\cite{SUS-11-028} require also two or more \cPqb-tagged jets.
In the first analysis ($\emu \ge 2\cPqb+\ETslash$), the \PWp, \PWm, and
\ttbar background distributions from simulation
are corrected to match the measured \ETslash spectrum at low \HT,
and then the corrected prediction is extrapolated to
high \HT and high \ETslash.
A selection on \ETslash significance (\ymet) and \HT is used
in the second analysis ($\emu \ge 3\cPqb$, \ymet).
The \ymet variable is defined as the ratio of \ETslash to $\sqrt{\HT}$.

\item[Opposite-Sign Dileptons]   Events are selected with two
leptons (electrons or muons) having electric charge of the opposite sign (OS), jets, and significant \ETslash.
In one (OS $\emu +\ETslash$) analysis~\cite{SUS-11-011},
a signal is defined as an excess of events at large values of \ETslash and \HT.
In a second (OS \emu edge) analysis~\cite{SUS-11-011}, a search is performed for a characteristic kinematic edge
in the dilepton mass distribution $m_{\ell^+\ell^-}$.
In these two analyses events with an $\Pe^+\Pe^-$~or~$\Pgm^+~\Pgm^-$~pair with invariant mass of the dilepton system between 76\GeV and 106\GeV
or below 12\GeV are removed, in order to suppress $\cPZ/\gamma^{*}$ events, as well as low-mass dilepton resonances.
A third analysis (OS \emu ANN)~\cite{SUS-11-018} applies a selection on the output of an ANN that is
based on seven kinematic variables constructed from leptons and jets,
to discriminate the signal events from the background.

Two other analyses complementary focus directly on the two leptons from \Z-boson decay
by applying an invariant mass selection~\cite{SUS-11-021}.
With this requirement, the main source of \ETslash arises from
fluctuations in the measurement of jet energy.
One analysis ($\cPZ+\ETslash$) determines this background
from a control sample that differs only in the presence of a \Z boson.
A second analysis (JZB) applies a kinematic variable denoted JZB, which is defined as
the difference between the sum of the vector elements of the \pt of the jets and the \pt of the boson candidate.

The last analysis in this group selects events consistent with a \PW~boson decaying to jets produced in association with a
\Z boson decaying to leptons, and searches for an excess of events in the \ETslash distribution.
This analysis is part of the combined lepton (comb. leptons) analysis~\cite{SUS-12-006} that targets
a signal containing gauge boson pairs and \ETslash: $\PW\cPZ$, $\cPZ\cPZ+\ETslash$.

\item[Same-Sign Dileptons]
Events are selected with two
leptons (electrons or muons) having electric charge of the same sign (SS), and significant \ETslash.
One analysis (SS \emu) uses several different selections on \ETslash and \HT
to suppress backgrounds~\cite{SUS-11-010}.
A second analysis (SS+\cPqb) requires at least one \cPqb-tagged jet~\cite{SUS-11-020}.
A third analysis makes no requirements on jet activity.
It limits backgrounds by applying more stringent lepton identification criteria.
Results from this analysis are included in the combined lepton results~\cite{SUS-12-006}.

\item[Multileptons]
Events are selected containing at least three leptons.
Selections are made on the values of several event variables, including \ETslash, \HT,
and the invariant mass of lepton pairs~\cite{SUS-11-013, SUS-12-006}.
One analysis applies a veto on \cPqb-tagged jets to remove most of
the \ttbar background, and is included in the combined lepton results~\cite{SUS-12-006}.

\item[Photons]
Events are selected containing one photon, two jets, and \ETslash ($\gamma \xspace \mathrm{jj}+\ETslash$), or
two photons, one jet, and \ETslash ($\gamma \gamma \xspace \mathrm{j} + \ETslash)$~\cite{SUS-12-001}.
The requirement of a photon and \ETslash is sufficient to remove most backgrounds.

\item[Inclusive]
The \textit{razor} analysis integrates several event categories~\cite{SUS-11-024}.
Events are required to contain jets and zero, one, or
two leptons (electrons and muons)
with a further classification based on the presence of a \cPqb-tagged jet.
The razor variable~\cite{Rogan:2010kb} is a ratio of a jet system mass to a transverse mass.
The distribution in the razor variable is highly correlated with the mass values of new particles
for hypothesized signals but skewed to relatively smaller values for backgrounds.
Values of the razor variable are chosen to reduce backgrounds while accepting signal events
in a similar manner as for the \AlphaT and \MTtwo analyses.
The razor analysis uses data corresponding to an integrated luminosity of $4.73\pm 0.10\fbinv$.

\end{description}

\section{Simplified models}
\label{sec:simplifiedmodels}

A simplified model is defined by a set of hypothetical
particles and a sequence of their production and decay.
In this paper, the selection of models is motivated by the particles and
interactions of the CMSSM or models with generalized gauge mediation~\cite{GGMe}.
For convenience, the particle naming convention of
the CMSSM is adopted, but none of the specific assumptions of the CMSSM are imposed.
The CMSSM assumptions include relationships among the new particle
masses, their production cross sections and distributions, and their
decay modes and distributions.
In the simplified models under consideration, only the production process for two primary particles is
considered.  Each primary particle can undergo a direct decay or a cascade decay through an intermediate
new particle.
Each particle decay chain ends with a neutral, undetected particle,
denoted LSP (lightest supersymmetric particle) in text and $\chioz$ in equations.
$\chioz$ can represent a neutralino or gravitino LSP.
The masses of the primary particle and the LSP are free parameters.
When the model includes the cascade decay of
a mother particle ($\text{mother}$) to an intermediate particle ($\text{int}$),
the mass of the intermediate particle depends on $m_{\text{mother}}$, $m_{\mathrm{LSP}}$, and a parameter $x$, according to the
equation \mbox{$m_{\text{int}} = x \, m_{\text{mother}} + (1-x) \, m_{\mathrm{LSP}}$}.
The value of $x$ can be anywhere in the range from zero to one, but values of
$x=\frac{1}{4},\frac{1}{2}$ and $\frac{3}{4}$ are used here.

The simplified models with a T1-, T3-, and T5-prefix are all models of gluino pair production,
with different assumptions about the gluino decay.
Those with a T2- and T6-prefix are models of squark-antisquark production, with different
assumptions on the type of squark or the pattern of squark decay.
Those with a TChi-prefix are models of chargino and neutralino production and decay.
In the simplified models under consideration, the \PW$\,$and \Z
bosons decay to any allowed final state.
A detailed description of the specific models follows.
Table~\ref{tab:allsmses} provides a summary.

\begin{description}

\item[T1, T1bbbb, T1tttt]
The T1 model is a simplified
version of gluino pair production.
Each gluino undergoes a three-body decay to a light-flavor
quark-antiquark pair and the LSP ($\gl \to \Pq\Paq \chioz$).
Ignoring the effects of additional radiation and jet reconstruction,
this choice produces a final state of 4 jets+\ETslash.
The T1bbbb and T1tttt models are modifications of the T1 model
in which the gluino decays
exclusively into \cPqb~or \cPqt~quark-antiquark pairs.
After accounting for the unobserved LSPs, the
kinematic properties of events from the T1tttt model are indistinguishable by the analyses considered here
to those from alternative simplified models, such as gluino
decay to a top squark and an anti-top quark, followed by
top squark decay to a top quark and
the LSP ($\gl\to\cPqt\sTop^*$, $\sTop^*\to\cPaqt\chioz$)
~\cite{SUS-11-020}.  For simplicity, only the T1tttt model is considered.

\item[T2, T2bb, T2tt, T6ttww]
The T2 model is a simplified
version of squark-antisquark production.  Each
squark
undergoes a two-body decay to a light-flavor quark and the LSP ($\sq \to \Pq \chioz$).
Ignoring the effects of additional radiation and jet reconstruction,
this choice produces a final state of 2 jets+\ETslash.
The T2bb and T2tt models are versions of bottom and top squark production, respectively,
with the bottom (top) squark decaying to a bottom (top) quark and the LSP.
The T6ttww model is a version of direct bottom squark production, with the bottom squark decaying to
a top quark, a \PW\ boson, and the LSP.

\item[T3w, T3lh]
The models with the T3-prefix are also based on gluino pair production.
One gluino has a direct decay to a light-flavor quark-antiquark pair and
the LSP, as in the T1 model.
The other gluino has a cascade decay through an intermediate particle, denoted as
$\chitz$ or $\chioc$.
In the T3w model, the cascade decay is a two-body decay of the
chargino to a \PW\ boson and the LSP.
For the T3lh model, the cascade decay is a three-body decay of a
heavy neutralino to a lepton pair and the LSP ($\chitz\to\ell^+\ell^-\chioz$).
If a heavy neutralino $\chitz$ decays to the LSP $\chioz$ and a pair of leptons,
the edge occurs at $m_{\ell^+\ell^-}=m_{\chitz}-m_{\chioz}$, corresponding
to the region of kinematic phase space where $\chioz$ is produced at rest in the $\chitz$ rest frame.

\item[T5lnu, T5zz]
The models with T5-prefix are also based on gluino pair production.
Both gluinos undergo cascade decays.   The T5lnu model has each
gluino decay to a quark-antiquark pair and chargino that undergoes a three-body decay to a lepton,
neutrino, and LSP.   The decay can produce SS dileptons, due to the
Majorana nature of the gluino.   The T5zz model has each gluino decay to a quark-antiquark pair
and an intermediate neutralino
that undergoes a two-body decay to a \Z boson and the LSP.
When both \Z bosons in an event
decay to a quark-antiquark pair, and
ignoring the effects of additional radiation and jet reconstruction,
the T5zz model produces a final state of 8 jets+\ETslash.

\item[TChiSlepSlep, TChiwz, TChizz]
These models are simplified versions of the direct
production and decay of charginos and neutralinos
or neutralino pairs.
The TChiSlepSlep and TChiwz models are versions of chargino-neutralino production.
The former has neutralino and chargino cascade decays through a charged slepton to three electrons, muons, and taus in equal rate,
while the latter has direct decays to gauge bosons and LSPs.
The TChiSlepSlep model does not include the decay $\chitz\rightarrow\PSgn\nu$, since this will not produce a multilepton signature.
The TChizz model, instead, is a version of neutralino pair production and decay into \Z bosons.

\item[T5gg, T5wg]
The T5gg model is a version of gluino pair production in which each gluino decays to
a quark-antiquark pair and an intermediate neutralino, which further decays to a photon and a massless LSP.
The T5wg model, instead, has one gluino decaying to quark-antiquark pair and an intermediate neutralino that decays
to a photon and the LSP, and the second gluino decaying to a quark-antiquark pair and a chargino that
decays to a \PW~boson and the LSP.   The neutralino and chargino masses are set to a common
value to allow an interpretation in models of gauge mediation.
The intermediate neutralino is labeled as the next-to-LSP (NLSP).

\end{description}

\begin{table*}[bpht]
\topcaption{Summary of the simplified models used in the interpretation of results.
\label{tab:allsmses}}
\centering
\begin{scotch}{rcccl}
\tablespace
        Model & Production & Decay  & Visibility & References \\
        name  & mode       &        &            &            \\   \hline
T1 & \gl \gl & {\small $\gl \rightarrow \qqbar\chioz$}& All-Hadronic &\cite{SUS-11-022,SUS-12-011,SUS-12-002,SUS-11-024}  \tablespace \\
T2 & \sq \sq$^*$ & {\small $\sq \rightarrow \Pq\chioz$}  & All-Hadronic &\cite{SUS-11-022,SUS-12-011,SUS-11-024} \tablespace  \\   \tablespace

T5zz & \gl \gl & {\small $\gl \rightarrow \qqbar \chitz$, $\chitz \rightarrow \Z \chioz$}  & All-Hadronic &\cite{SUS-12-011,SUS-12-002} \\
         &            &                                  & Opposite-Sign Dileptons    &\cite{SUS-11-021}   \\ \tablespace
         &            &                                  & Multileptons    &\cite{SUS-11-013}  \\     \tablespace
T3w & \gl \gl & {\small $\gl \rightarrow \qqbar\chioz$}  & Single Lepton + Jets &\cite{SUS-12-010} \\  \tablespace
        &    & {\small $\gl \rightarrow \Pq \Paq \chipm$, $\chipm \rightarrow \Wpm \chioz$} & &  \tablespace \\    \tablespace

    T5lnu & \gl \gl & {\small $\gl \rightarrow \Pq \Paq \chipm$, $ \chipm  \rightarrow \ell \nu \chioz$}  & Same-Sign Dileptons &\cite{SUS-11-010} \tablespace \\    \tablespace
    T3lh & \gl \gl & {\small $\gl \rightarrow \Pq \Paq \chioz$}   & Opposite-Sign Dileptons &\cite{SUS-11-011,SUS-11-018} \\ \tablespace
         & & {\small $\gl \rightarrow \Pq \Paq \chitz $, $\chitz \rightarrow \ell^+ \ell^- \chioz$}  & &  \tablespace \\     \tablespace
    T1bbbb & \gl \gl & {\small$\gl \rightarrow \bbbar\chioz$}  & All-Hadronic~(b)  &\cite{SUS-12-002,SUS-12-003,SUS-11-022,SUS-11-024} \tablespace \\    \tablespace

    T1tttt & \gl \gl & {\small$\gl \rightarrow \ttbar\chioz$}  & All-Hadronic~(b)  &\cite{SUS-12-003,SUS-11-022,SUS-12-002} \\
           &            &                                        & Single Lepton + Jets~(b)   &\cite{SUS-11-028}  \\ \tablespace
           &            &                                        & Same-Sign Dileptons~(b)   &\cite{SUS-11-010,SUS-11-020}  \\ \tablespace
           &            &                                        & Inclusive~(b)   &\cite{SUS-11-024}  \\    \tablespace

    T2bb & $\sBot ~\sBot^*$ & {\small $\sBot \rightarrow \cPqb\chioz$}  & All-Hadronic~(b) &\cite{SUS-11-022,SUS-11-024}  \tablespace \\    \tablespace
    T6ttww & $\sBot ~\sBot^*$ & {\small $\sBot \rightarrow \cPqt\chim$, $\chim \to \PWm \chioz$}  & Same-Sign Dileptons~(b) &\cite{SUS-11-020}  \tablespace \\    \tablespace
    T2tt & $\sTop~\sTop^*$ & {\small $\sTop \rightarrow \cPqt\chioz$}  & All-Hadronic~(b) &\cite{SUS-11-022,SUS-11-024}  \tablespace \\    \tablespace

    {\small TChiSlepSlep} & $\chipm \chitz$ & {\small $\chitz \rightarrow \ell^\pm \tilde{\ell}^\mp$, $\tilde{\ell} \rightarrow \ell \chioz$}  & Multileptons &\cite{SUS-11-013,SUS-12-006} \\ \tablespace
                          &                & {\small $\chipm \rightarrow \nu \tilde{\ell}^{\pm}$, $\tilde{\ell}^{\pm} \rightarrow \ell^{\pm} \chioz$} &   & \tablespace \\     \tablespace

    {\small TChiwz} & $\chipm \chitz$  & {\small $\chipm \rightarrow \Wpm \chioz$, $\chitz \rightarrow \Z \chioz$}  & Multileptons &\cite{SUS-11-013,SUS-12-006}  \tablespace \\    \tablespace

    {\small TChizz} & $\chitz \chithreez$  & {\small $\chitz, \chithreez \rightarrow Z \chioz$}  & Multileptons &\cite{SUS-11-013,SUS-12-006} \tablespace \\    \tablespace

T5gg & $\gl~\gl$ & $\gl\to \qqbar\chitz$, $\chitz\to\gamma\chioz$    & Photons &\cite{SUS-12-001} \tablespace \\    \tablespace
T5wg & $\gl~\gl$ & $\gl\to \qqbar\chitz$, $\chitz\to\gamma\chioz$    & Photons &\cite{SUS-12-001} \\ \tablespace
     &          & $\gl\to \Pq\Paq\chipm$, $\chipm\to \Wpm\chioz$    &         &                   \\
\end{scotch}
\end{table*}

The calculation of \accXeff for each simplified model uses
the \PYTHIA~\cite{Pythia} event generator with
the SUSY differential cross sections for gluino, squark-antisquark, and neutralino and chargino pair production.
The decays of non-Standard Model particles are performed with a constant amplitude, so that
no spin correlations exist between the decay products.
The primary particle masses are varied between 100\GeV and 1500\GeV.
The theoretical prediction for the
production cross section is not needed
to calculate \sigmaXBFUL.  However, it is informative to compare the values
of \sigmaXBFUL with the production cross section expected in a benchmark model.
The selected benchmark
is the CMSSM cross section prediction for
gluino pair, squark antisquark,
or neutralino and chargino  pair production.
The cross sections are determined at next-to-leading order (\SigNLO) accuracy for gaugino pair
production, and at NLO with next-to-leading-logarithmic contributions (\SigNLL) for the other processes~\cite{Beenakker:1996ch,Kulesza:2008jb,Kulesza:2009kq,Beenakker:2009ha,Beenakker:2011fu,Kramer:2012bx}.
In both the calculation of particle production and the reference cross sections,
extraneous SUSY particles are decoupled.  For example, the contributions from
squarks are effectively removed by setting the squark mass to a very
large value when calculating the acceptance and cross section for gluino
pair production.
The cross sections are presented under the assumption of unit branching ratios, even for models such as T3w that consider two different decay modes of the gluino.

\section{Limit setting procedure}
\label{sec:procedure}

The method used to set exclusion limits is common to all simplified models and analyses.
In this section, the procedure is presented using the
interpretation of two different OS dilepton analyses~\cite{SUS-11-011} as an example.

The reference simplified model is T3lh, which can yield
pairs of OS leptons not arising from \Z-boson decays.
The mass of the intermediate neutralino produced in the gluino decay chain is set using
$m_{\chitz}=\frac{1}{2}( m_{\gl} + m_{\chioz} )$, corresponding to $x=\frac{1}{2}$.
The parameter $x$ influences the patterns of cascade decays.
The mass splitting between the gluino and the intermediate particle, $(1-x)(m_{\gl} - m_{\chioz})$,
influences the observable hadronic energy, while the
mass splitting between the intermediate particle and the LSP, $x \xspace ( m_{\gl} - m_{\chioz})$,
influences the energy of the leptonic decay products or the \ETslash.
For large $x$, the signal is expected to have lower \HT and higher
\ETslash, and possibly higher-\pt leptons.  Conversely, for small $x$, the
signal should have higher \HT and lower \ETslash, and possibly
lower-\pt leptons.
Results are shown for a counting experiment based on \HT and \ETslash selections (OS \emu + \ETslash), and
the edge reconstruction in the dilepton invariant mass ($m_{\ell^+\ell^-}$) distribution (OS \emu edge).

\begin{figure*}[ht]
\begin{center}
  \includegraphics[width=0.45\textwidth]{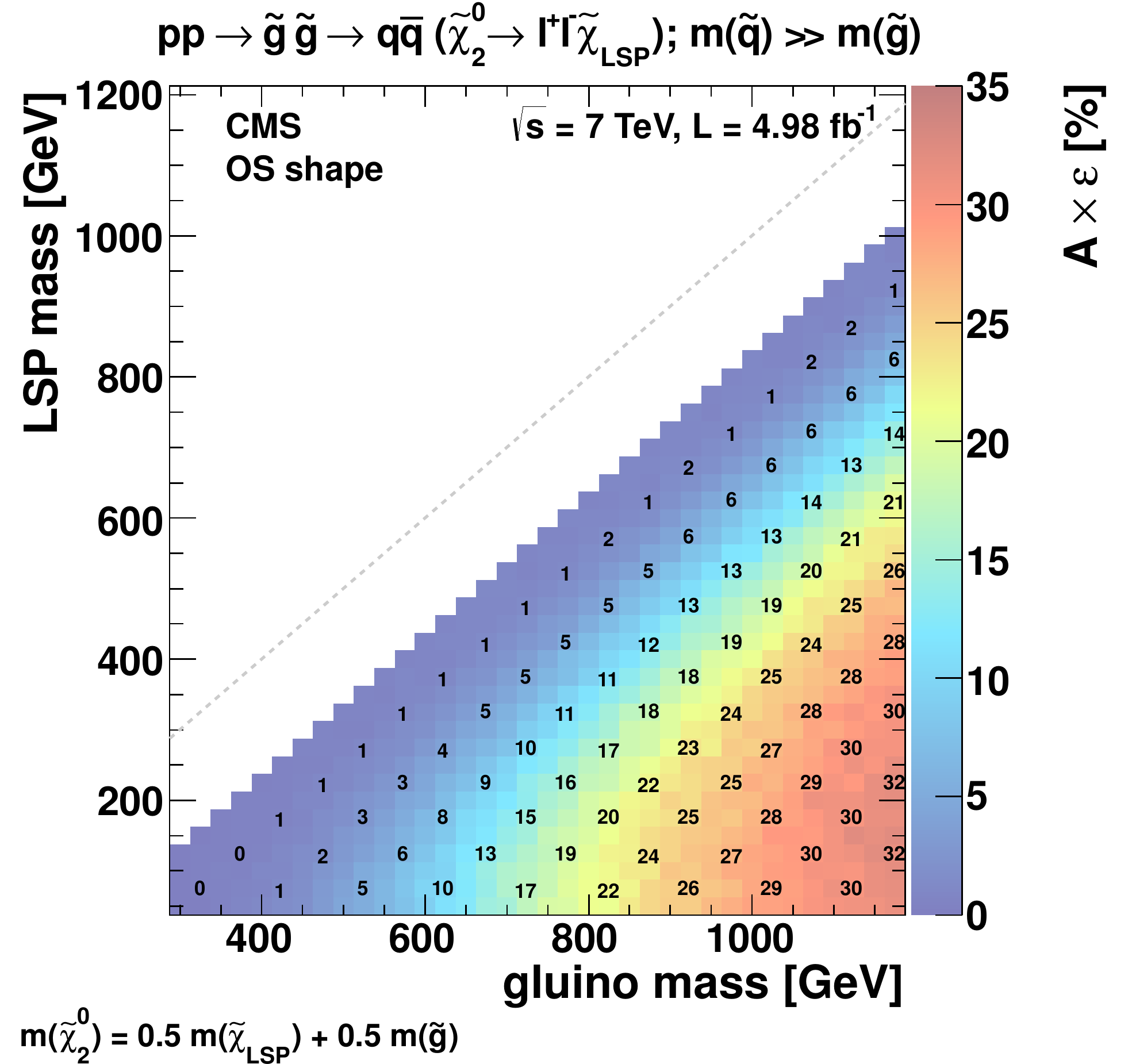}
  \includegraphics[width=0.45\textwidth]{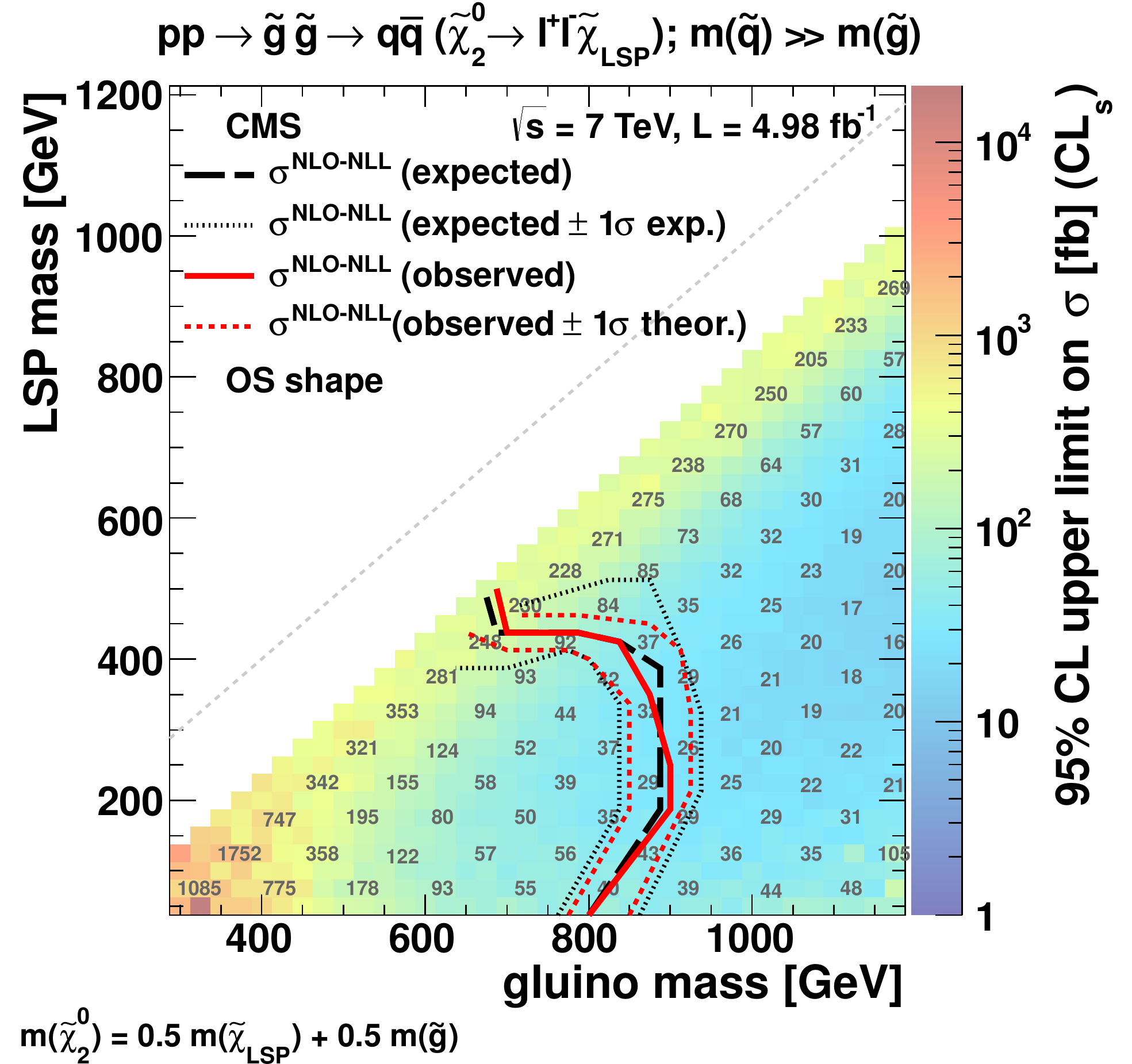}
  \includegraphics[width=0.45\textwidth]{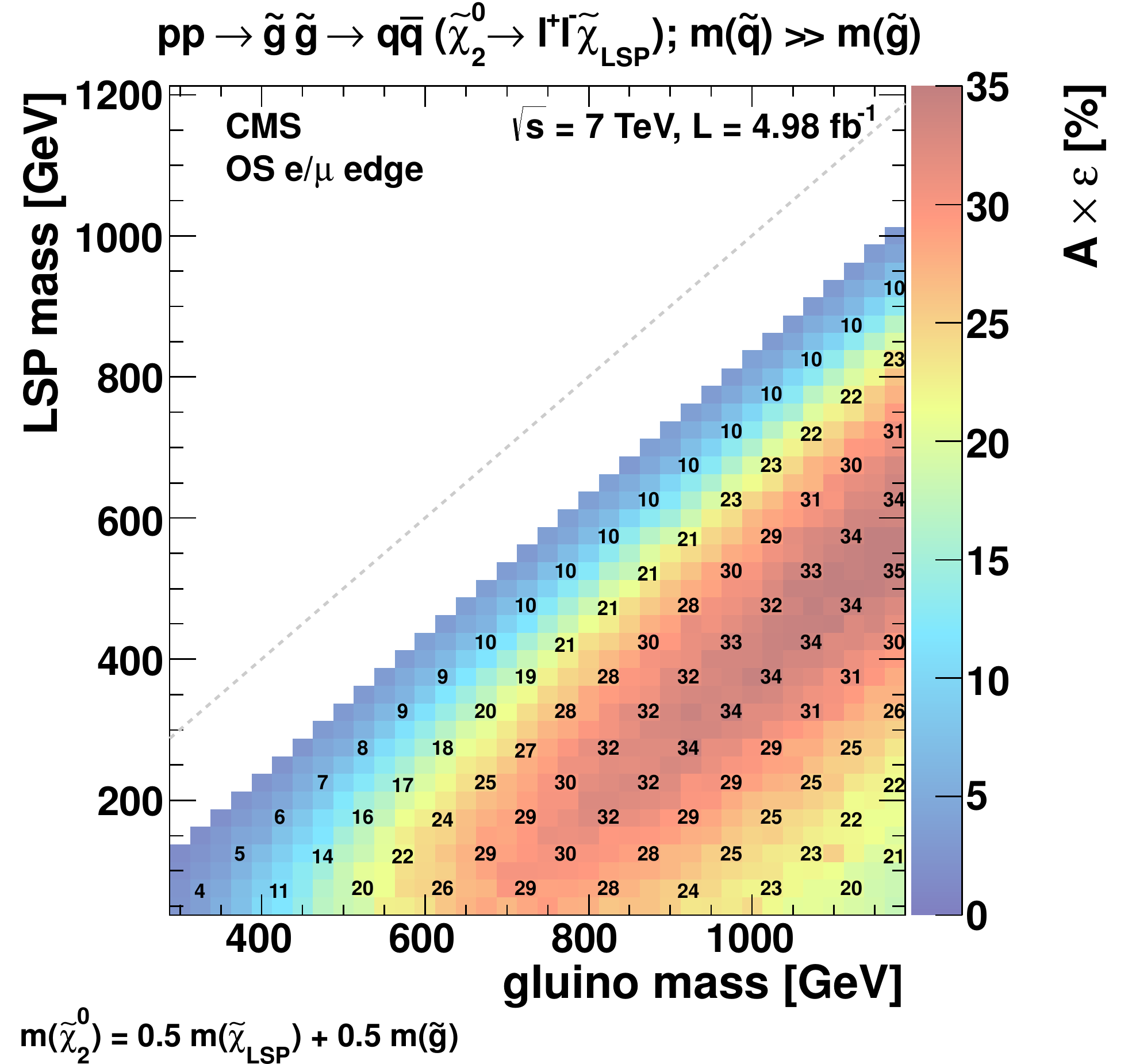}
  \includegraphics[width=0.45\textwidth]{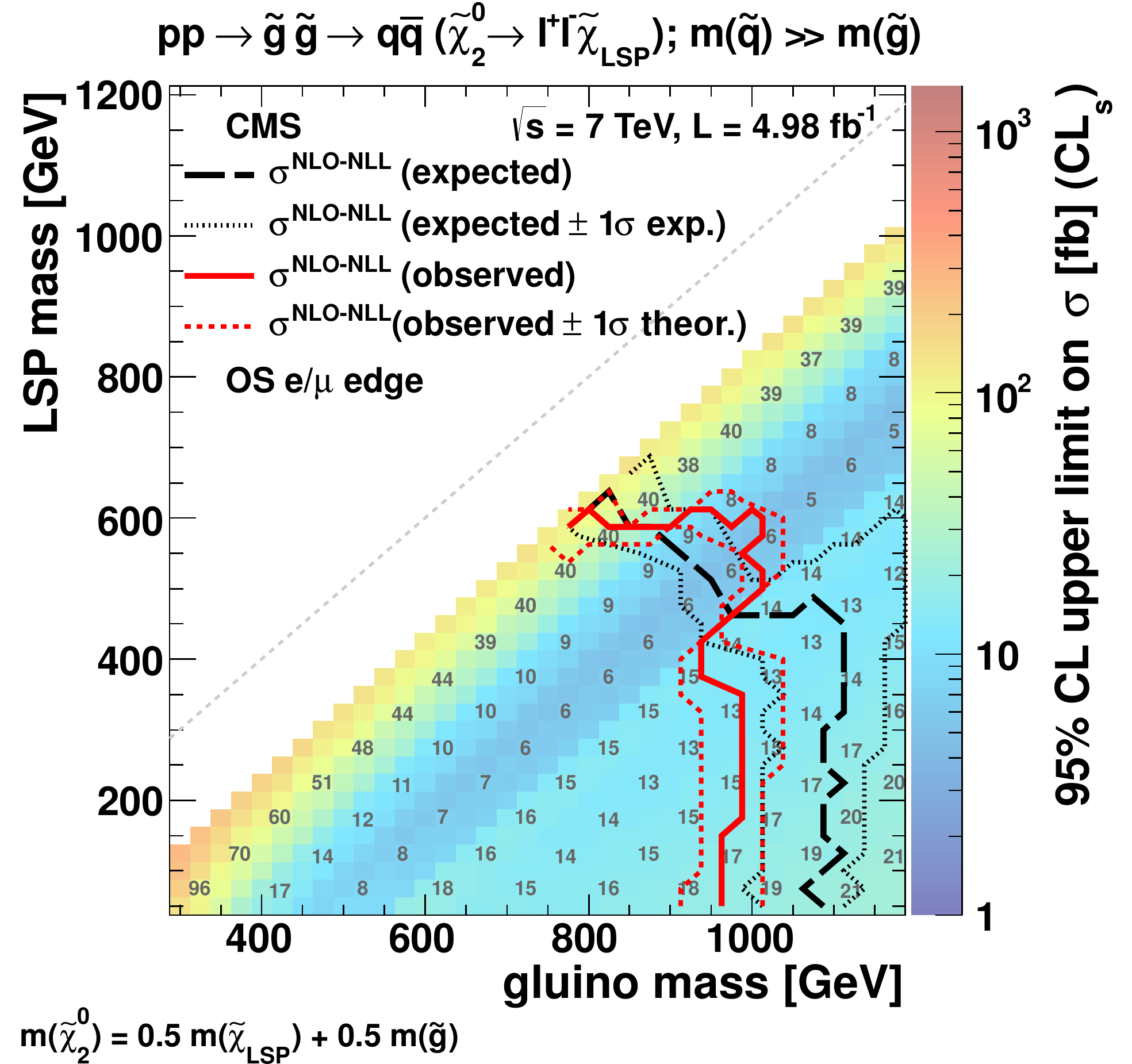}
\caption{OS dileptons~\cite{SUS-11-011}: Product of the experimental efficiency and acceptance (left)
and the upper limit on the product of the cross section and branching fraction (right) for the T3lh model
from the \ETslash and \HT selection (top)
and from edge reconstruction (bottom).
Results are shown as a function
of gluino and LSP mass, with the intermediate neutralino mass set using $x=0.5$.
}
\label{fig:dileptonResultsOS}
\label{fig:T1lhOS_eff}
\label{fig:T1lhOS_limit}
\end{center}
\end{figure*}

In the first step, the event selection is applied to simulated simplified model events.
The ratio of selected to generated events determines $\accXeff$.
The uncertainty on this quantity, which is a necessary input to the limit calculation,
is described in the analysis references.
This calculation is repeated for
different values of the gluino and LSP masses.
Values of $\accXeff$ for the two selections are shown in Figure~\ref{fig:T1lhOS_limit} (left).

The acceptance of the \ETslash-based analysis (top-left) increases with increasing gluino mass,
since the \ETslash and gluino mass are correlated.
However, this acceptance decreases for smaller gluino-LSP mass splitting.
The acceptance of the edge-based analysis (bottom-left) is relatively larger
for small gluino-LSP mass splitting, but
decreases for larger gluino mass.
This decrease is an artifact of a choice made in the analysis to limit the $m_{\ell^+\ell^-}$
distribution to $m_{\ell^+\ell^-}<300\GeV$.
For small mass splitting, the presence of initial-state radiation (ISR) can strongly
influence the experimental acceptance.   The uncertainty on ISR is difficult to
estimate.   For this reason, some analyses report results for only a restricted
region of mass splittings.

Estimates of \accXeff, the background, and their uncertainties are used to calculate \sigmaXBFUL
for the given model using the $\mathrm{CL}_\mathrm{s}$ criterion~\cite{Read:2002hq,Junk:1999kv}.
A gluino and LSP mass pair in a simplified model is excluded if
the derived \sigmaXBFUL result is below the predicted \SigNLL for those mass values.

Values of \sigmaXBFUL are
shown in Figure~\ref{fig:T1lhOS_limit} (right) for the two analyses.
The edge analysis (bottom-right) has less a stringent selection on \ETslash than the counting experiment (top-right):
for the former the signal regions are defined by $\HT>300\GeV$ and $\ETslash>150\GeV$,
 while for the latter different signal regions are obtained requiring high \HT
($\HT>600\GeV$ and $\ETslash>200\GeV$), high \ETslash  ($\HT>300\GeV$ and $\ETslash>275\GeV$)
or tight selection criteria ($\HT>600\GeV$ and $\ETslash>275\GeV$).
As a result, the edge analysis sets stronger limits than the counting analysis in this particular topology.

The expected limit and its experimental uncertainty, together with the observed limit
and its theoretical uncertainty based on \SigNLL
for gluino pair production, are shown as curves overlaying the exclusion limit.
In the case of the OS \emu edge analysis, the most stringent limit on the gluino mass is obtained at around 900\GeV for low LSP masses and for the OS \emu + \ETslash and is at
around 775\GeV. The quoted estimates are determined from the observed exclusion based on the theoretical production cross section minus $1\sigma$ uncertainty.

The contours of constant \accXeff do not coincide with those of constant \sigmaXBFUL. For the OS \emu + \ETslash, this is an artifact of the
changing uncertainty on \accXeff as a function of gluino and LSP masses, signal contamination, and the observed distribution of \ETslash.
In the OS \emu edge analysis, this occurs because the allowed area of the signal $m_{\ell^+\ell^-}$ distribution varies with the mass difference between the gluino and LSP.

\section{Results and comparisons}
\label{sec:comparisons}

This Section presents the results obtained applying the
procedure described in Section~\ref{sec:procedure} to the CMS analyses
presented in Section~\ref{sec:analyses}.
The individual results are described in detail in the analysis references,
but comparisons of the results are presented in this paper for the first time.
For each analysis, the lower limit on
particle masses in a simplified model is determined
by comparing \sigmaXBFUL with the predicted \SigNLL or \SigNLO as described in Section~\ref{sec:simplifiedmodels}.
Only the observed \sigmaXBFUL values are used.  The limits are thus
subject to statistical fluctuations.

\begin{figure*}[htp]
\begin{center}
  \includegraphics[width=0.8\textwidth]{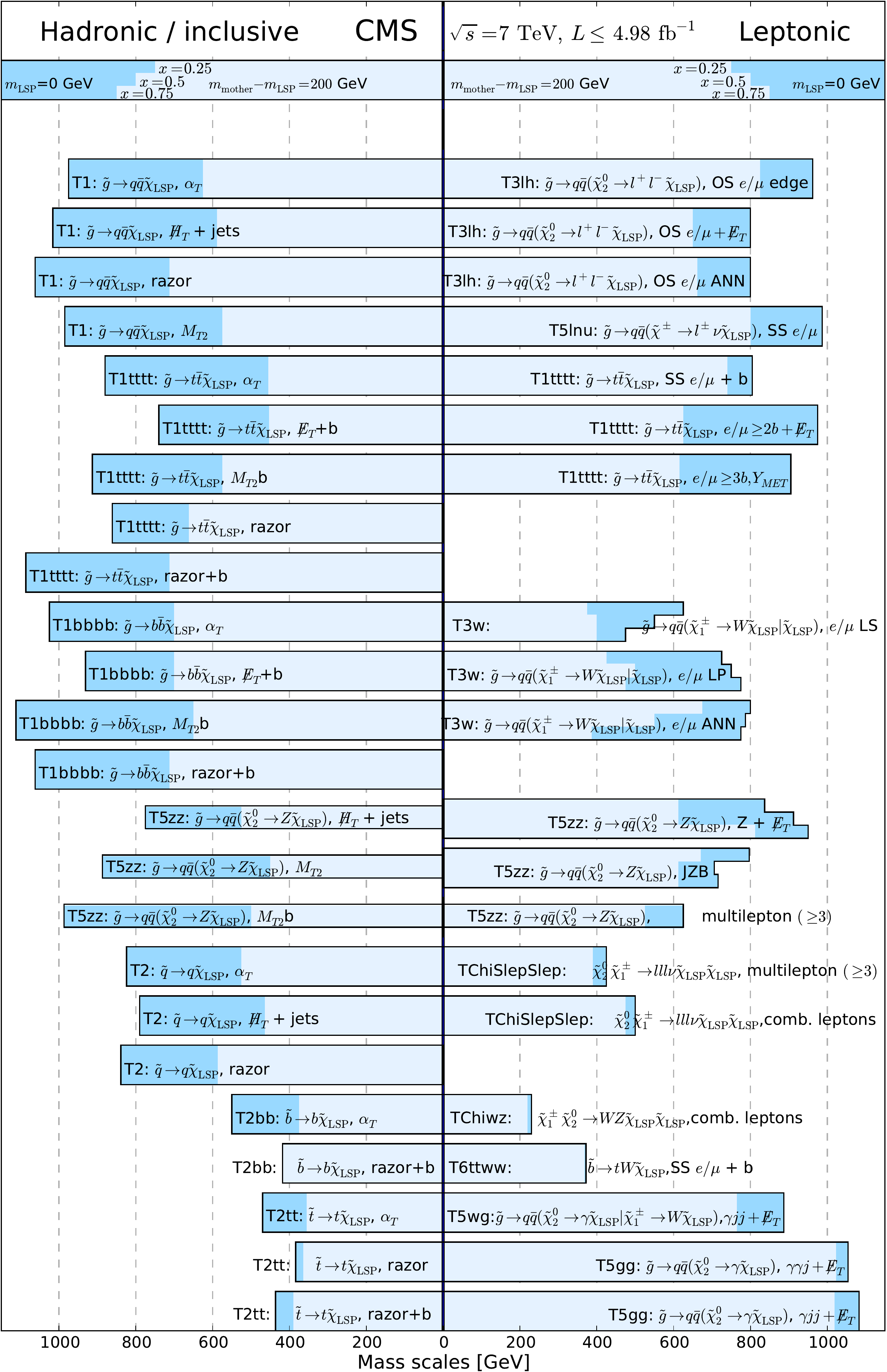}
\end{center}
   \caption{Exclusion limits for the masses of the mother particles, for
$m_\mathrm{LSP}=0\GeV$ (dark blue) and $m_\text{mother}-m_\mathrm{LSP}=200\GeV$
 (light blue), for each analysis, for the hadronic and razor results (left) and the leptonic results (right).
The limits are derived by comparing the allowed \sigmaXBFUL to the theory described in the text.
 For the T3, T5 and TChiSlepSlep models, the mass of the intermediate particle is defined by the relation
$m_{\text{int}} = x \, m_{\text{mother}} + (1-x) \, m_{\mathrm{LSP}}$.
 For the T3w and T5zz models, the results are presented for $x=0.25, 0.5, 0.75$, while for the T3lh, T5lnu, and TChiSlepSlep models,
 $x=0.5$.
  The lowest mass value for $m_\text{mother}$ depends on the particular analysis and the simplified model.
   \label{fig:barPlots}}
\end{figure*}

\begin{figure*}[htb]
   \begin{center}
   \includegraphics[width=0.8573\textwidth]{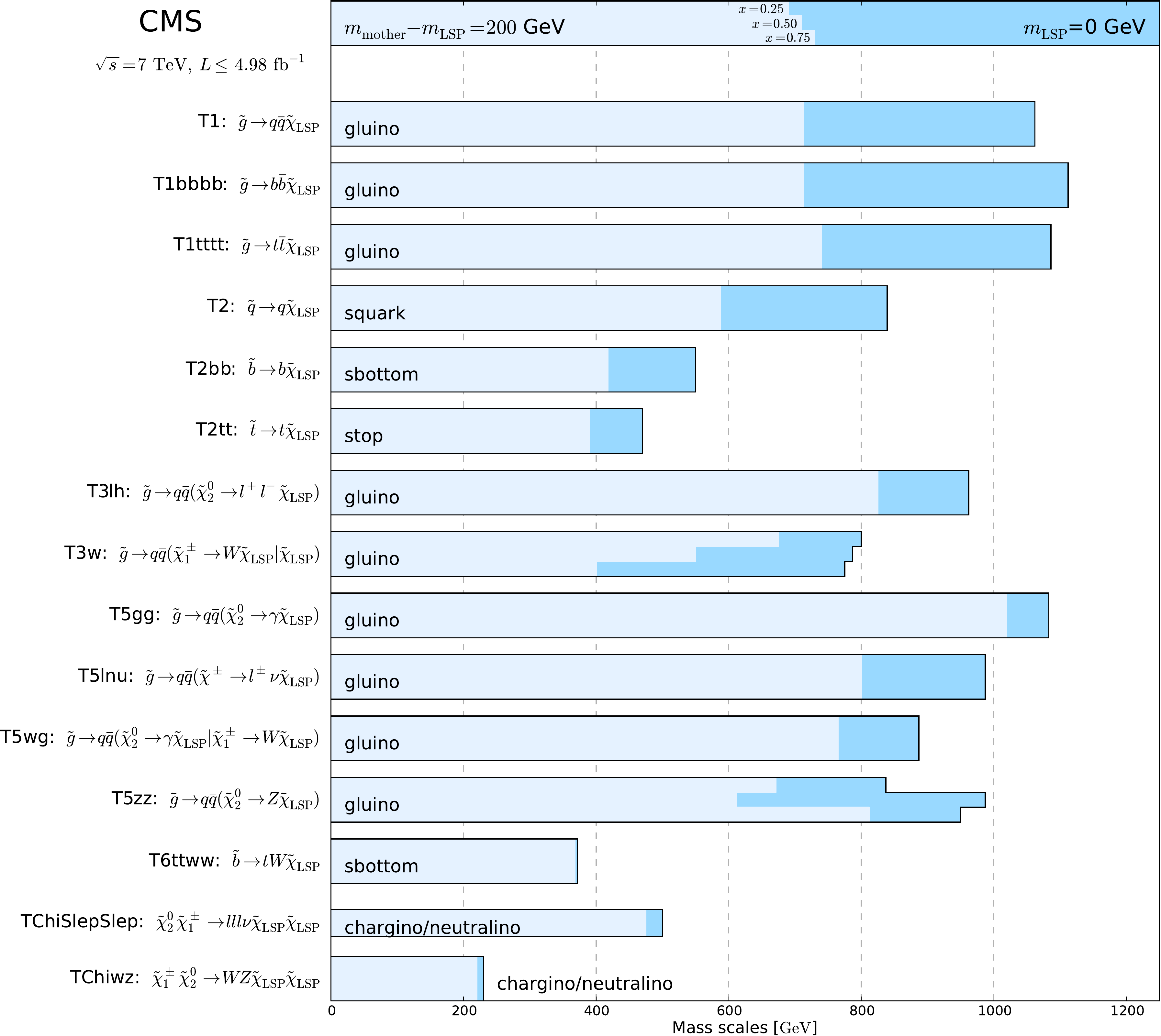}% scale relative to previous plot: 1.0716
   \caption{Best exclusion limits for the masses of the mother particles,
    for $m_\mathrm{LSP}=0\GeV$ (dark blue) and $m_\text{mother}-m_\mathrm{LSP}=200
\GeV$ (light blue), for each simplified model, for all analyses considered.
 For the T3, T5 and TChiSlepSlep models, the mass of the intermediate particle is defined by the relation
$m_{\text{int}} = x \, m_{\text{mother}} + (1-x) \, m_{\mathrm{LSP}}$.
 For the T3w and T5zz models, the results are presented for $x=0.25, 0.5, 0.75$, while for the T3lh, T5lnu, and TChiSlepSlep models,
 $x=0.5$.
The lowest mass value for $m_\text{mother}$ depends on the particular analysis and the simplified model.
   \label{fig:barPlotBrief}}
    \end{center}
\end{figure*}

Figure~\ref{fig:barPlots} illustrates the results of the hadronic and inclusive analyses (left) and the leptonic analyses (right).  Comparisons
are made for two reference points of the mother and LSP masses: one with a massless LSP ($M0$, dark blue
in Figure~\ref{fig:barPlots}), one with a fixed mass splitting between the mother particle and the LSP of 200\GeV ($\Delta M200$,
light blue in Figure~\ref{fig:barPlots}).
The results shown in Figure~\ref{fig:barPlots} are summarized below.

\begin{description}

\item[All-Hadronic]
This class of analyses sets limits on those models, such as T1, T2, and T5zz,
that produce several jets, but few leptons.
The \AlphaT and \HsTjets analyses yield similar limits in the T1 and T2
models despite the differences in their event selections.
In the case of the T5zz model, the \MTtwo analysis is more sensitive to the model's mass splitting than
the \HsTjets analysis:
for $M0$, the \MTtwo analysis sets the stronger limit, while for $\Delta M200$ the \HsTjets analysis
is more sensitive.  This is expected, since the \MTtwo analysis uses a higher cut on \HT than the \HsTjets analysis.
In general, the limits for the T5zz model are reduced with respect to the T1 and T2 models, because
of the reduced amount of \ETslash in cascade decays.

The \ETslash+\cPqb, the \MTtwob, and the \AlphaT $\xspace$ analyses set limits on the
T1bbbb, T1tttt, T2bb, and T2tt models,
visible in Figure~\ref{fig:barPlots} (left).  The three analyses set comparable limits
for $\Delta M200$, but the \MTtwob~and \AlphaT $\xspace$ analyses set the stronger limits for $M0$.
For the T1tttt model, the \MTtwob~analysis is most sensitive.

The \MTtwob~analysis is also compared with the \MTtwo analysis with no \cPqb-tagging requirement.
The limit for the \MTtwob~analysis on the T1bbbb model
is stronger than for the \MTtwo analysis on the T1 model, since many
of the backgrounds are removed by requiring a \cPqb-tagged jet, allowing for
a lower threshold on the \MTtwo variable.
Also, the limit on the T5zz model from the \MTtwob~analysis is stronger than for the
\MTtwo analysis,
even though the \cPqb-tagged jets from the T5zz model arise mainly through the decay $\Z\to\bbbar$.

\item[Single Lepton + Jets]
This class of analyses is sensitive to simplified models that produce \PW~ bosons or direct decays
to leptons.
The \emu LS, LP, and ANN analyses set limits on the T3w model
for an intermediate (chargino) mass corresponding to $x=\frac{1}{4},\frac{1}{2}$, and $\frac{3}{4}$.

The LS and LP analyses are sensitive to the kinematic properties of the \PW~ boson produced in the chargino decay.
For a large mass splitting between the mother and LSP ($M0$), the LP and ANN limits
are less sensitive to $x$ than the LS limit.  For a fixed mass splitting ($\Delta M200$), however,
the ANN limit is more sensitive.
The limits for $M0$ are stronger for all three analyses, with LP and ANN setting the best limits.

The $\emu \ge 2\cPqb+$\ETslash and $\emu \ge 3\cPqb$,~\ymet analyses set limits on the T1tttt model.

\item[Opposite-Sign Dileptons]
The Z+\ETslash and JZB analyses both set limits on the T5zz model relying on
the leptonic decays of one of the \Z bosons.
The Z+\ETslash analysis sets the stronger limit for
$x=\frac{3}{4}$ and $M0$, for which more \ETslash is produced on average.
The JZB analysis has the opposite behavior, since the separation between
signal and background in the JZB variable is maximized
in the signal when the \ETslash and Z-\pt vectors point in the same direction.
Therefore, the best limit is set for $x=\frac{1}{4}$.

Limits are also set on
the T3lh model, with the non-resonant decay of the intermediate neutralino
to leptons, by the \ETslash, the edge-based, and the neural-network-based analyses.
The edge-based analysis sets significantly stronger limits.

\item[Same-Sign Dileptons]
The T5lnu model produces equal numbers of OS and SS dileptons.
Limits are set on the T5lnu model by the SS dilepton analysis.
No comparisons are made for the OS dilepton analyses as
these are expected to be much less sensitive due to their larger backgrounds.

The SS dilepton analysis with a \cPqb-tagged jet is used to set limits on the T1tttt and the T6ttww models.
The analysis is not strongly sensitive to mass splittings, and
a similar limit is set for the case of $M0$ or $\Delta M200$.

\item[Multileptons]
Limits are set on the TChiSlepSlep model, which produces
leptons through slepton decays but not through gauge-boson decays.
A limit is set on the chargino mass (which equals the heavy neutralino mass) near
500\GeV, which is
not strongly dependent on the mass splitting.
The limits on the model TChiwz are significantly reduced because of
the corresponding reduction from the branching fraction of
the gauge bosons into leptons.
A limit on the T5zz model is also set.  For the $\Delta M200$
case the limit is competitive with the limits set by the hadronic analyses,
despite the low $\Z\rightarrow\ell^+\ell^-$ branching fraction.

\item[Photons]  Limits are set on the T5gg and T5wg models, which produce
two isolated photons and \ETslash or one isolated photon and \ETslash, respectively.
The one- and two-photon analyses set comparable limits on the T5gg model.
In addition, the one-photon analysis sets a competitive limit on the T5wg model.

\item[Inclusive] The razor and razor+\cPqb~analyses set limits on the T1, T2, T1bbbb,
T1tttt, T2bb, and T2tt models.  The limits on each of these models are comparable with the best limits set by
individual, exclusive analyses.

\end{description}

Figure \ref{fig:barPlotBrief} illustrates the best hadronic or leptonic result for each
simplified model.
Excluding the photon signatures,
the best limits for the $M0$ scenario exclude gluino masses below 1\TeV and squark masses
below 800\GeV.
For the $\Delta M200$ scenario, the limit is reduced to near 800\GeV and 600\GeV, respectively.
The limits on the gluino mass from the photon signatures are near 1.1\TeV, regardless
of the mass splitting.

\begin{figure*}[htbp]
\begin{center}
  \includegraphics[width=0.45\textwidth]{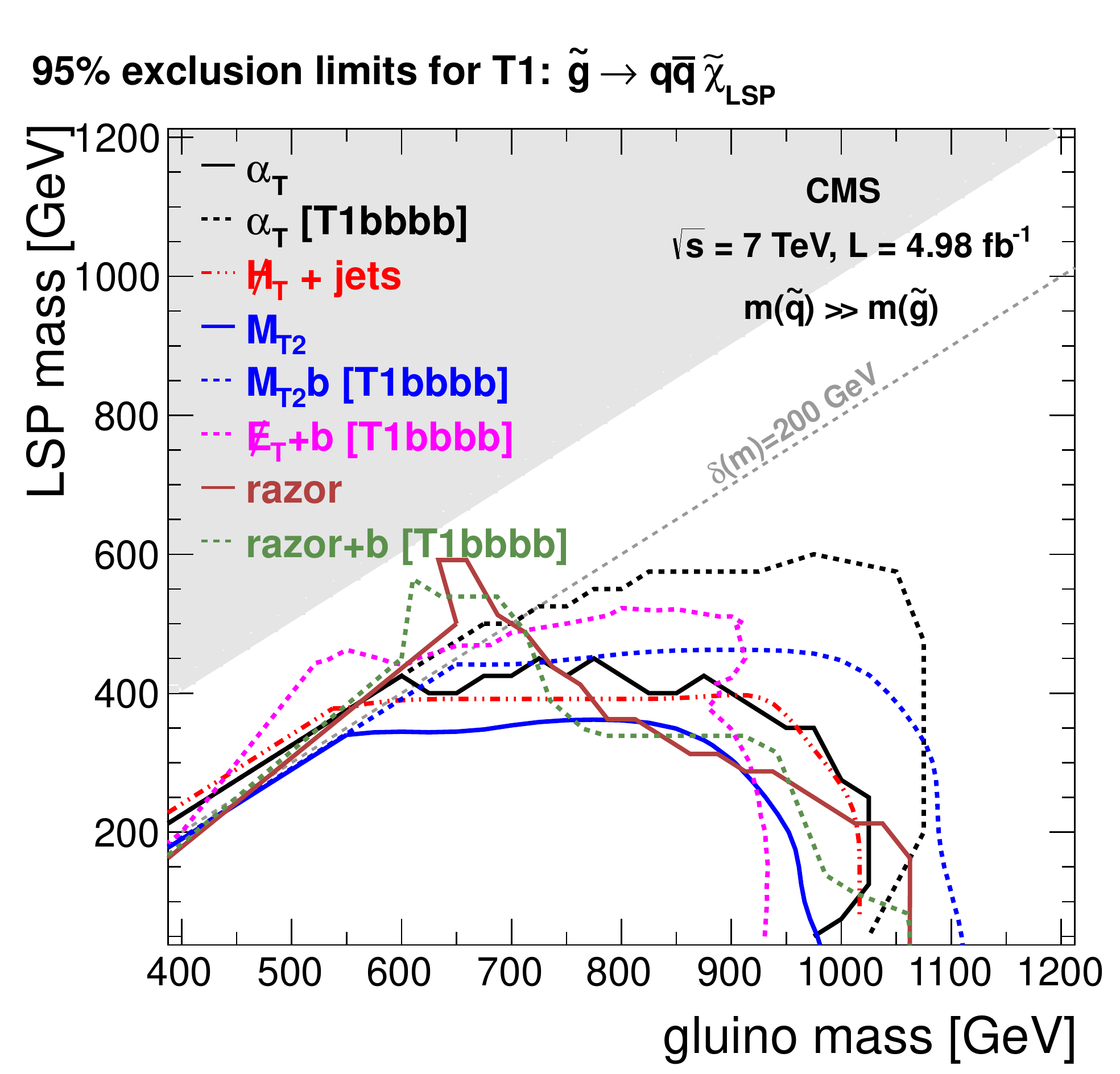}
  \includegraphics[width=0.45\textwidth]{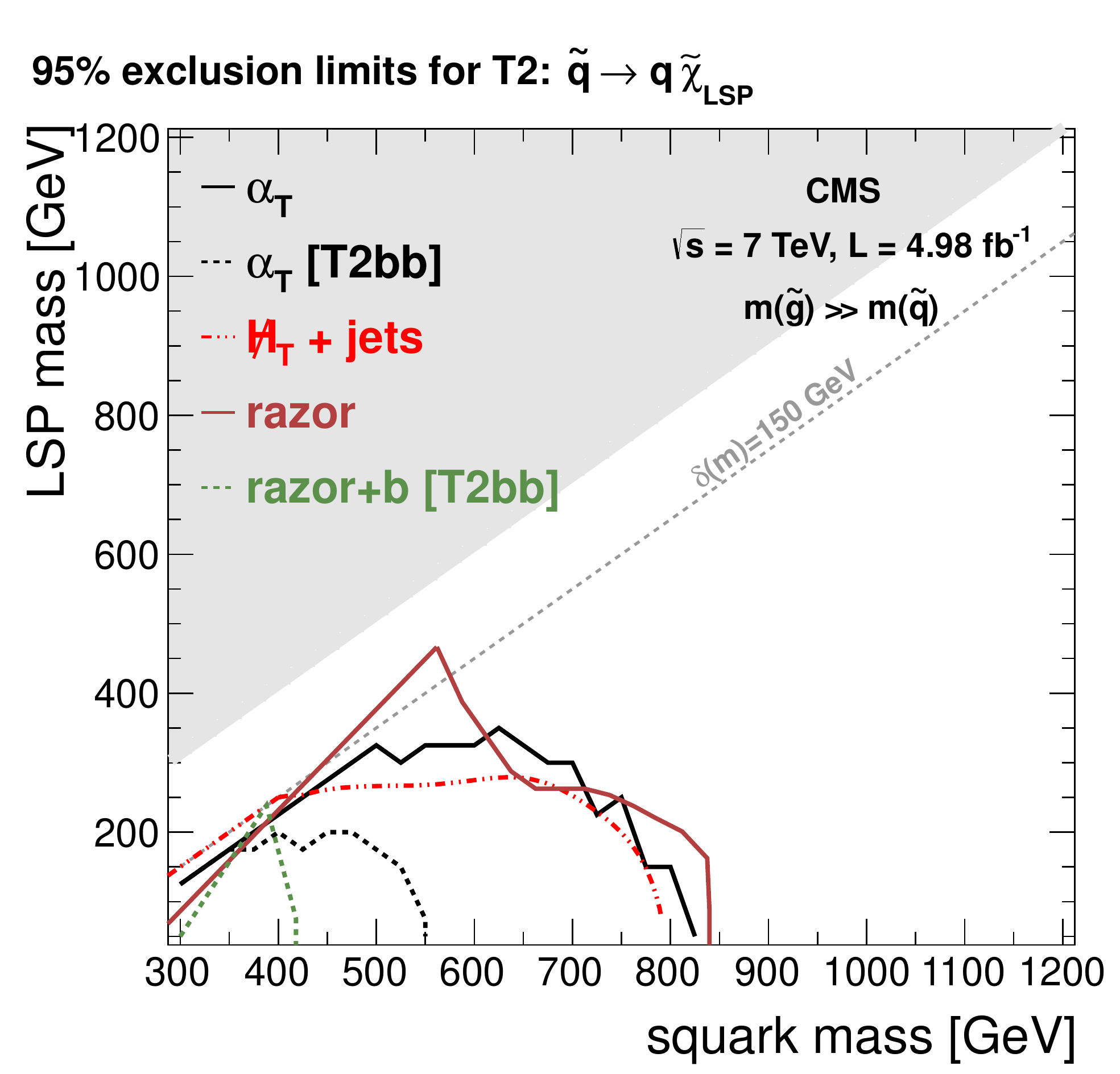}
  \includegraphics[width=0.45\textwidth]{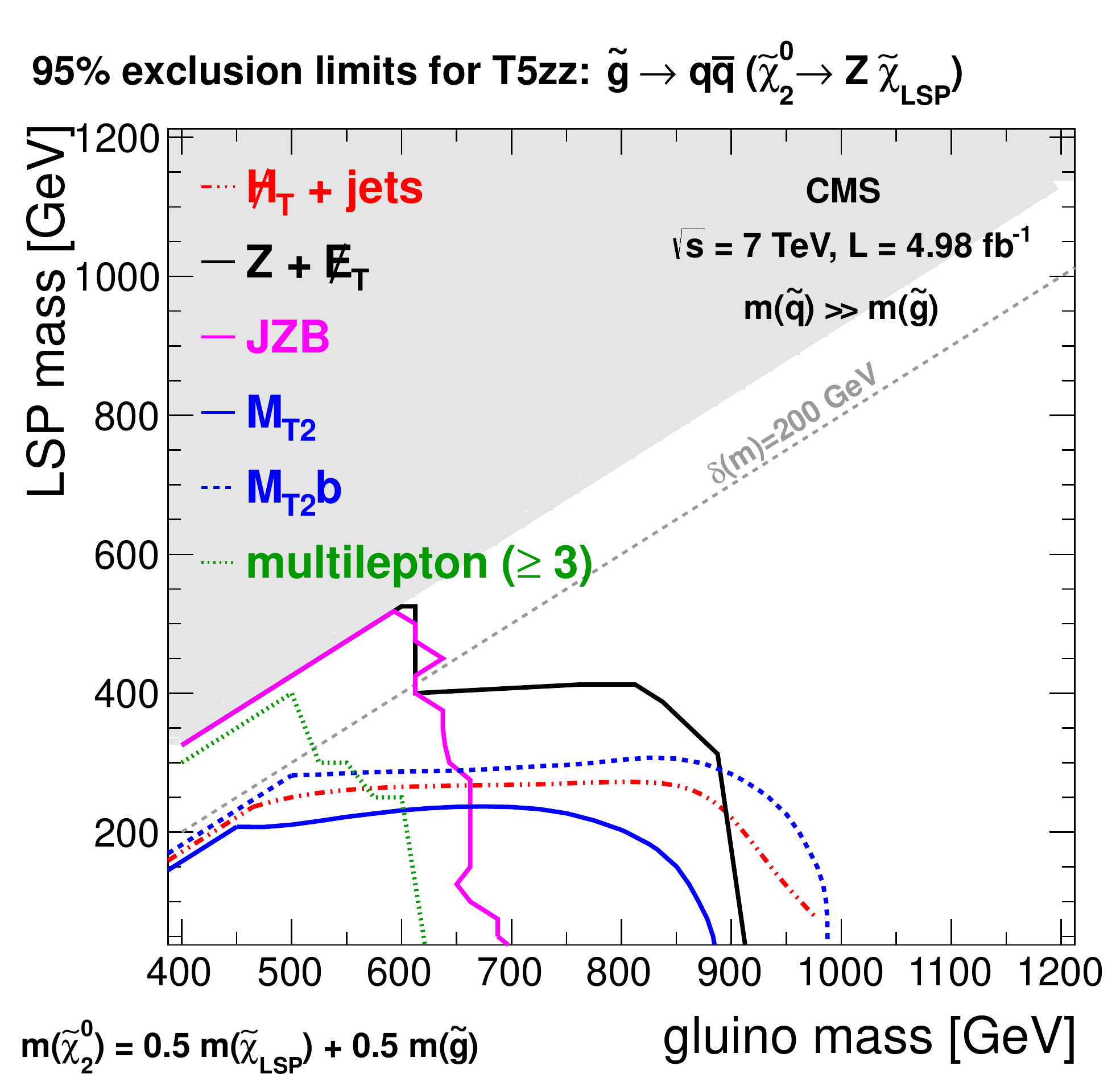}
  \includegraphics[width=0.45\textwidth]{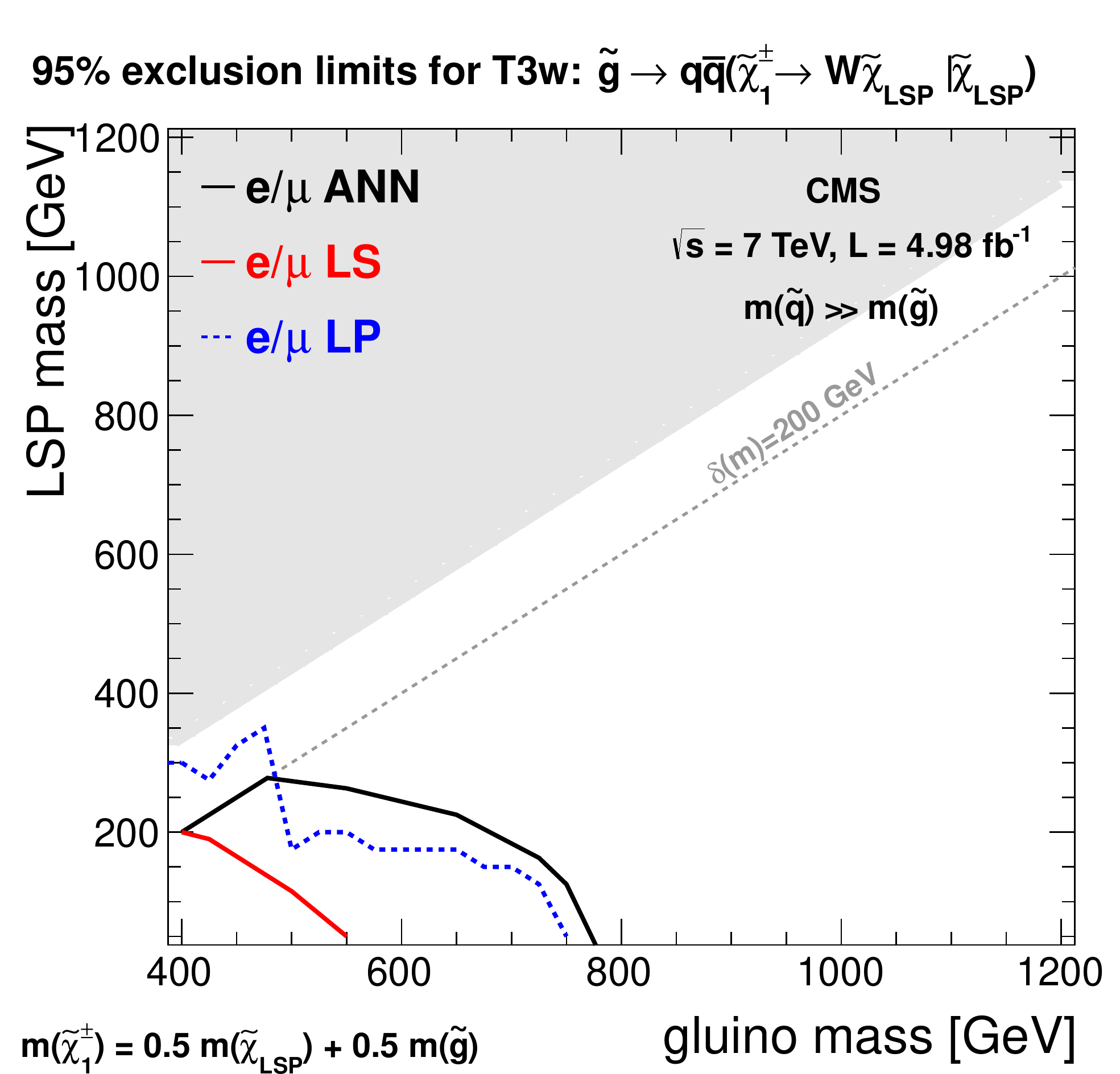}
  \includegraphics[width=0.45\textwidth]{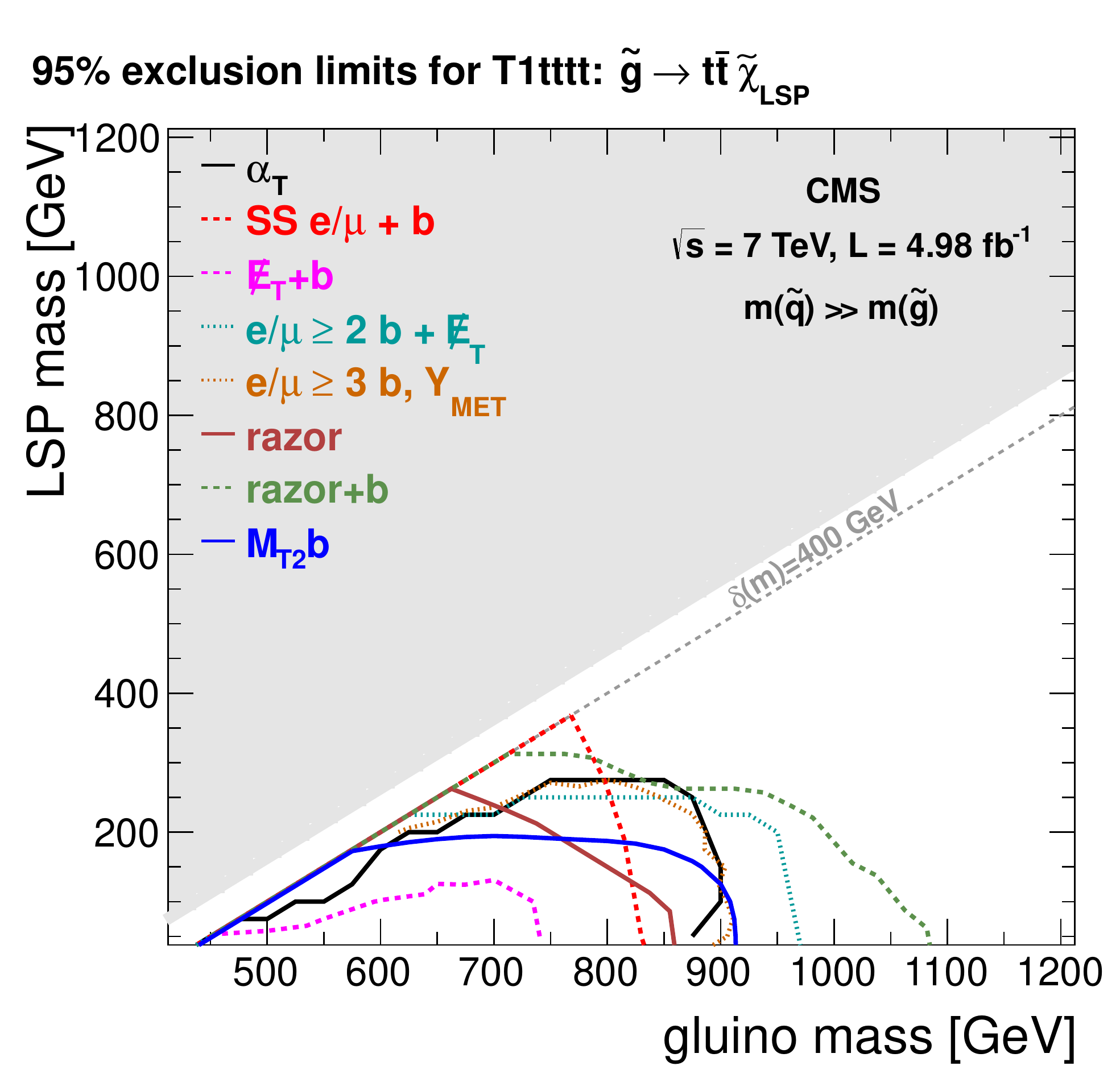}
  \includegraphics[width=0.45\textwidth]{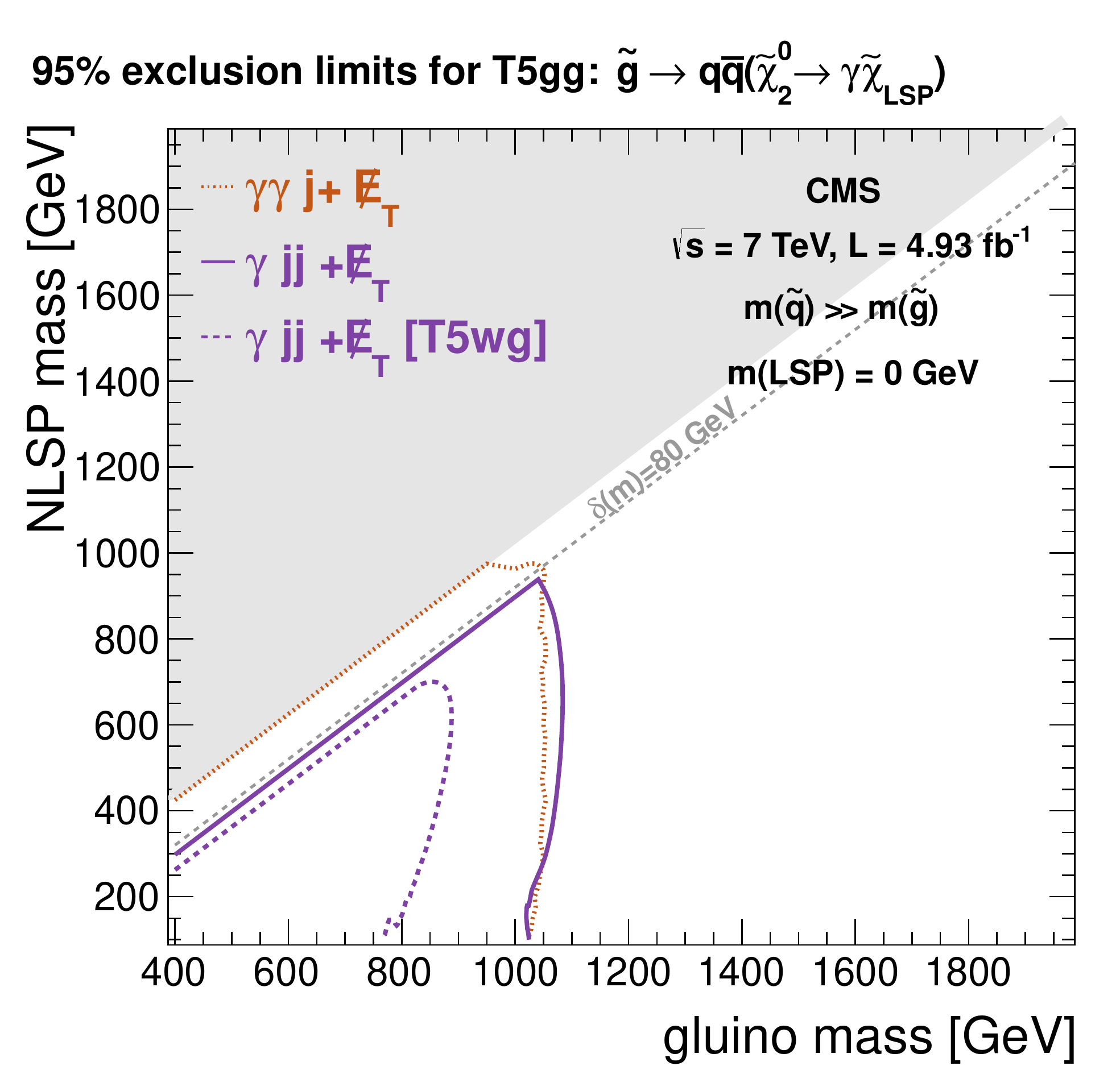}
\caption{
The 95\% CL exclusion limits on the produced particle and LSP masses
in the models T1(T1bbbb), T2(T2bb), T5zz, T3w, T1tttt, and
T5gg(T5wg). The grey area represents the region where the decay mode is forbidden.}
\label{fig:coveragePlots}
\end{center}
\end{figure*}

Figure~\ref{fig:coveragePlots} illustrates the exclusion contours in the two-dimensional
plane of the mother versus LSP mass for
the T1 (T1bbbb), T2 (T2bb), T5zz, T3w, T1tttt and T5gg (T5wg) models.
The results shown in Figures~\ref{fig:barPlots} and \ref{fig:barPlotBrief} are a subset of
these results.
Regions where
the analyses, due to the uncertainty in the acceptance
calculation, do not produce a limit are denoted by dashed lines.
Figure~\ref{fig:coveragePlots} (upper-left)
shows the exclusion contours of the T1 and T1bbbb models using the hadronic and \cPqb-tagged hadronic analyses.
This tests the dependence on the assumption of whether the gluino decays to
light or heavy flavors.
Solid (dashed) lines are used for the T1 (T1bbbb) model.
The \AlphaT analysis covers a larger area in the gluino-LSP
mass plane for the T1bbbb model than the hadronic decays do for the T1 model.
However, this comparison is only valid if the gluino indeed
decays only to bottom quarks.
The fully hadronic \HsTjets and \AlphaT $\xspace$ analyses
cover a similar region, while the \MTtwo analysis covers comparatively less.
The inclusive analysis is particularly sensitive when the difference in mass between the mother and LSP is small,
a situation known as a ``compressed spectrum.''

Figure~\ref{fig:coveragePlots} (upper-right)
compares the exclusion contours of the T2 and T2bb models.  The \AlphaT $\xspace$ and \HsTjets analyses set similar limits on the
T2 model.
The \AlphaT analysis sets weaker limits on the T2bb model, but the
reference cross section is a factor of eight smaller than for the T2 model.
The inclusive analysis sets the overall strongest limits, particularly in the low mass splitting region.

Figure~\ref{fig:coveragePlots} (middle-left) compares the exclusion contours of the T5zz model.
The T5zz model comparison demonstrates the complementarity of leptonic, hadronic,
and \cPqb-tagged hadronic analyses.
In particular, the leptonic analyses are more limiting for smaller mass splittings,
while the hadronic analyses are more limiting for larger gluino masses.

Figure~\ref{fig:coveragePlots} (middle-right) compares the exclusion contours of the T3w model.
The \emu ANN and \emu LP analyses
provide comparable results.  The \emu LS spectrum analysis excludes a
smaller region.

Figure~\ref{fig:coveragePlots} (bottom-left) compares the exclusion contours of the T1tttt model.
The inclusive analysis with b-tagged jets sets the strongest limit on the gluino mass.
The SS+\cPqb~analysis, however, sets limits that are almost independent of mass splitting.

Figure~\ref{fig:coveragePlots} (bottom-right) compares the exclusion contours of the
T5gg and T5wg models.  The limits on the T5gg and T5wg models
demonstrate the insensitivity of these photon analyses to the NLSP mass.
Also, the requirement on the number of photons (one or two) has little effect on the limit on the T5gg model.
The limit on the T5wg model, which has only one signal photon per event, excludes a
smaller region than the limit on the T5gg model.

\begin{figure*}[htbp]
\begin{center}
  \includegraphics[width=0.45\textwidth]{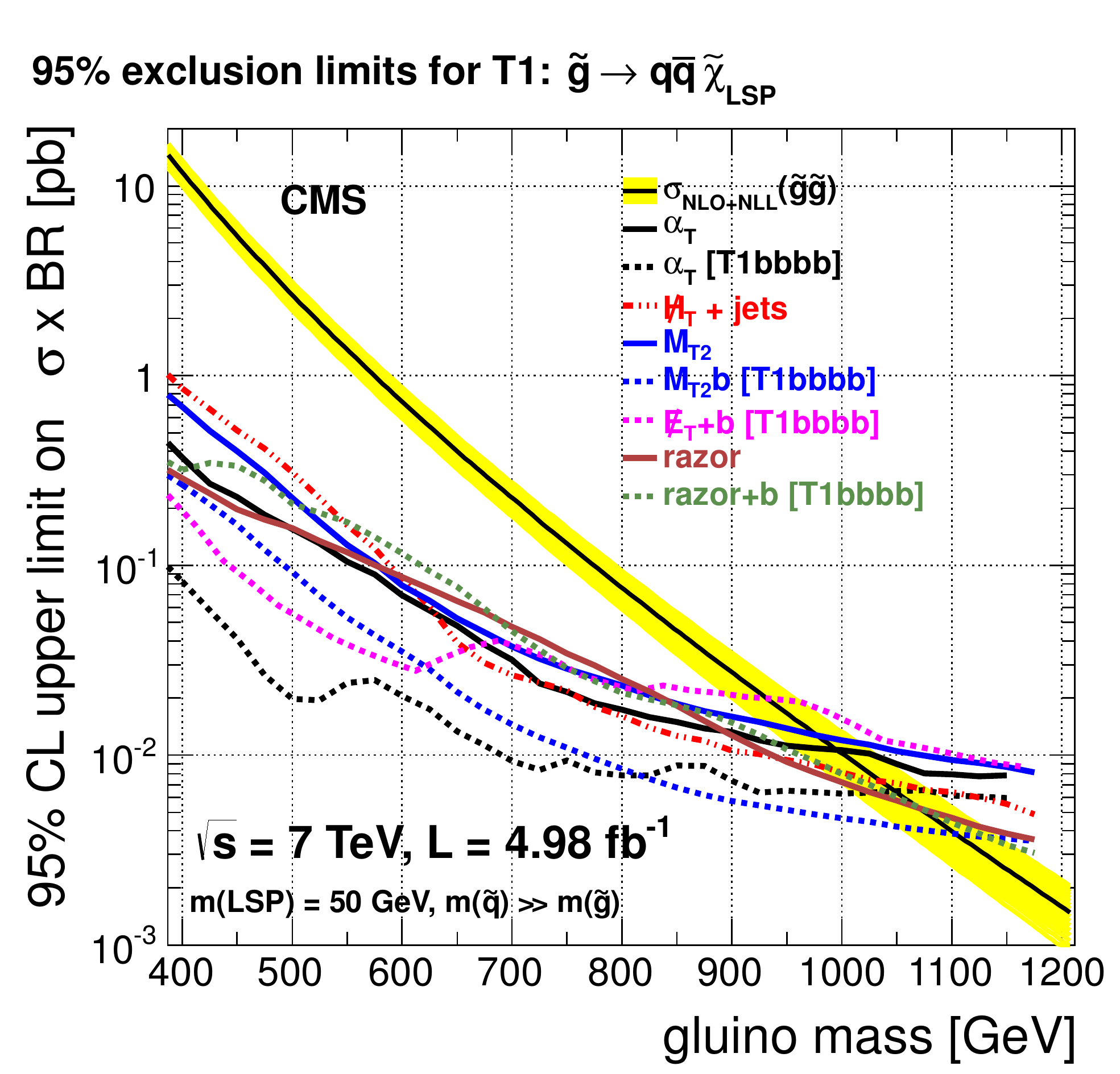}
  \includegraphics[width=0.45\textwidth]{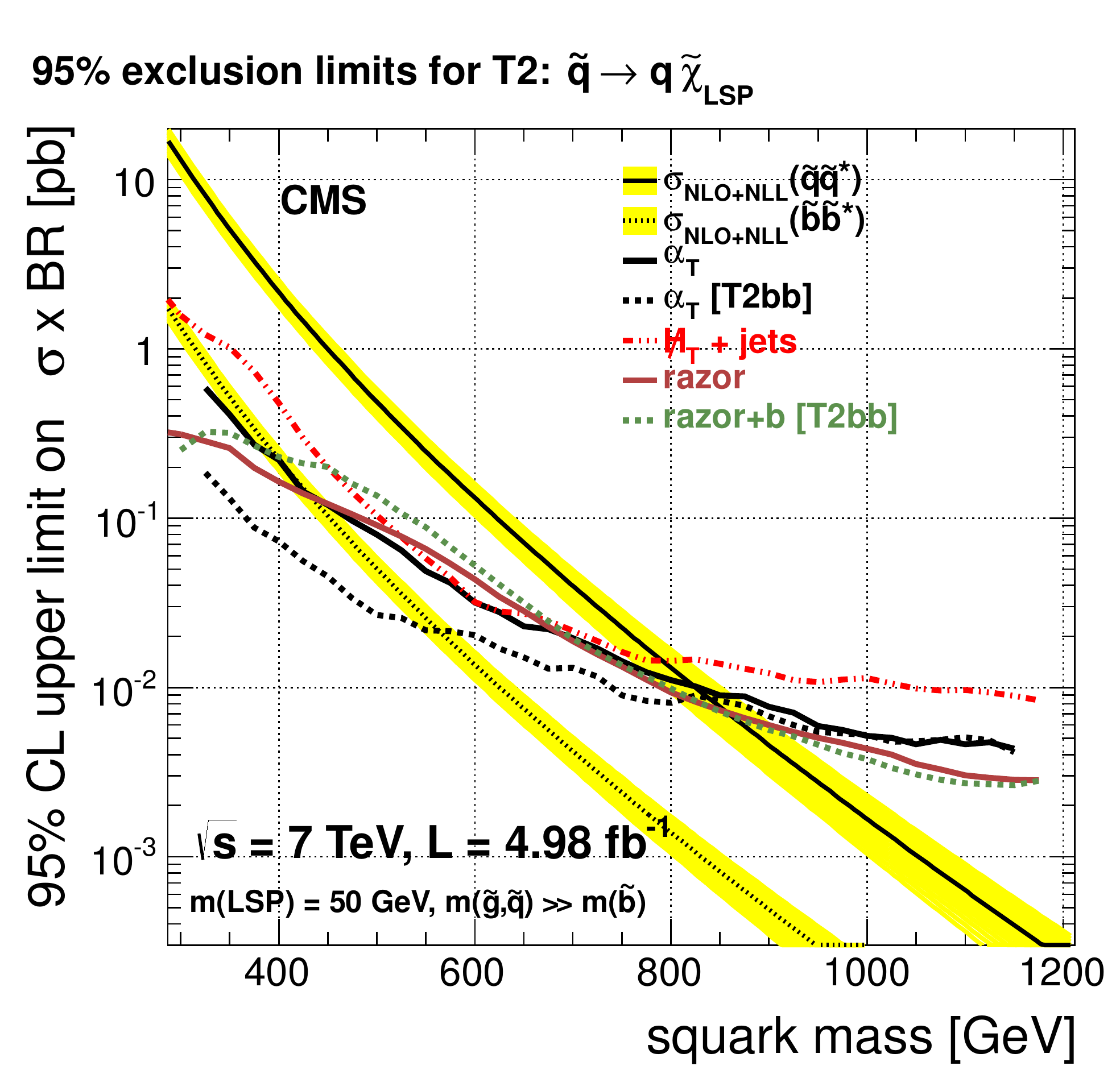}
  \includegraphics[width=0.45\textwidth]{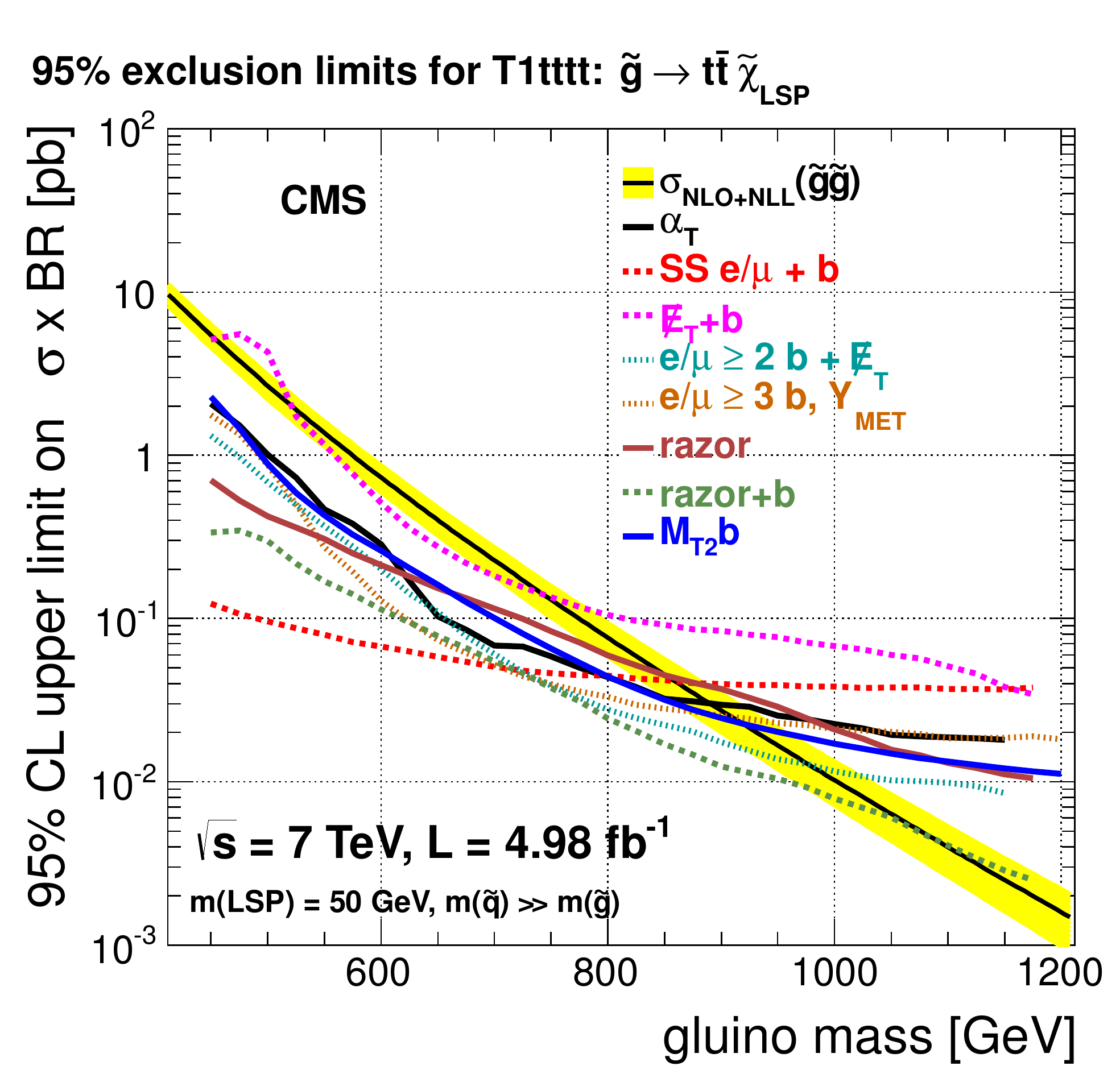}
  \includegraphics[width=0.45\textwidth]{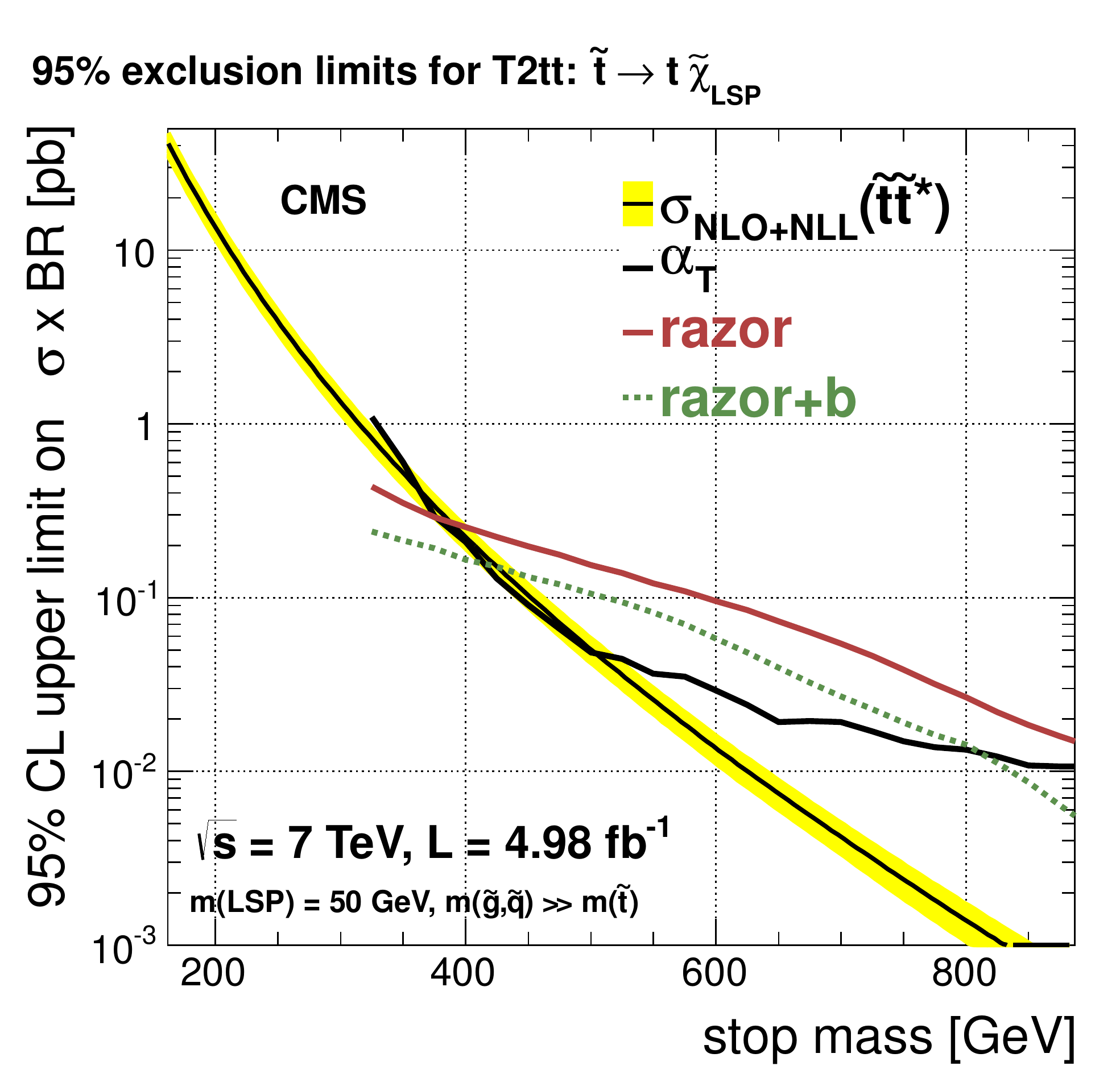}
  \includegraphics[width=0.45\textwidth]{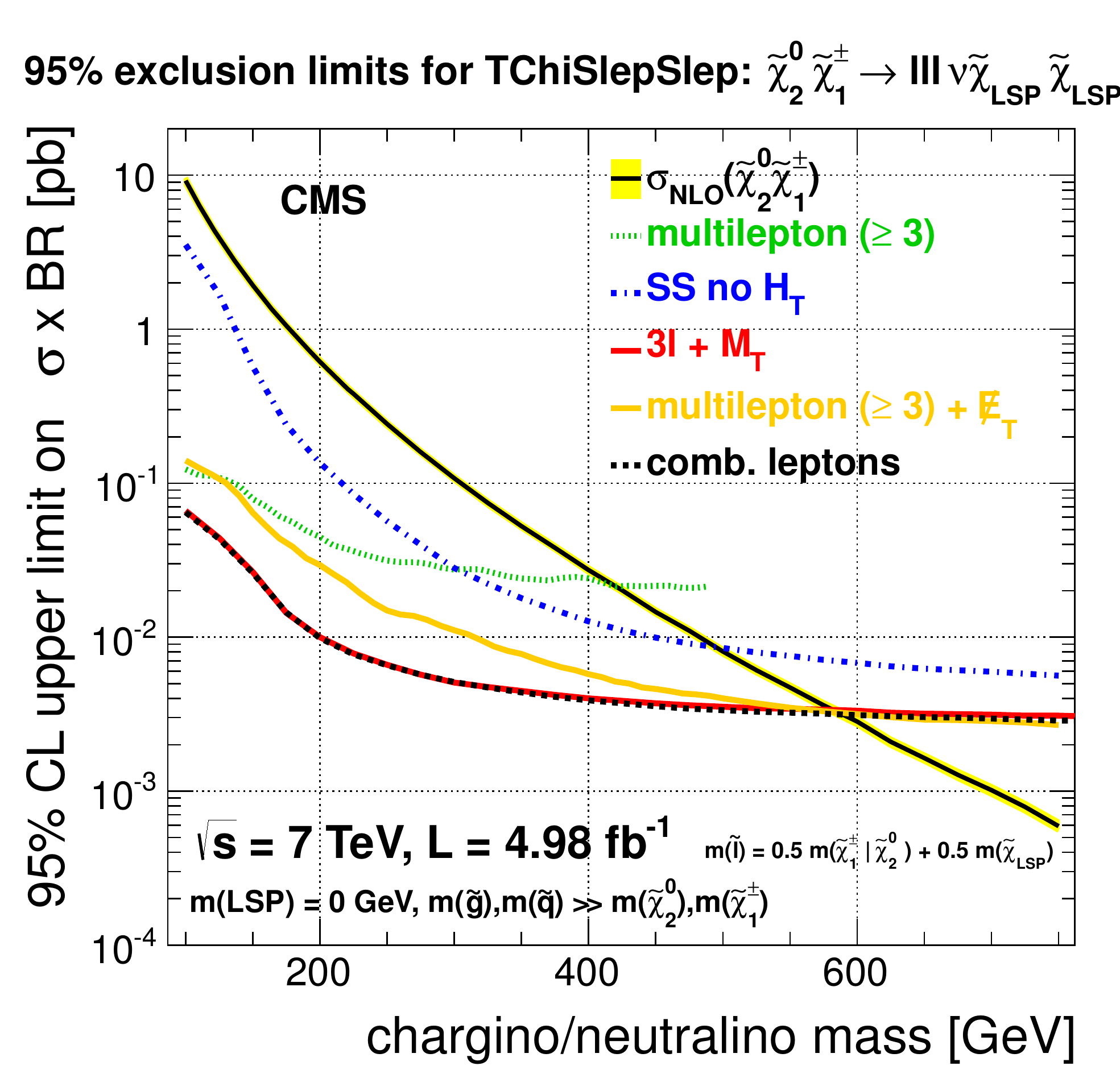}
  \includegraphics[width=0.45\textwidth]{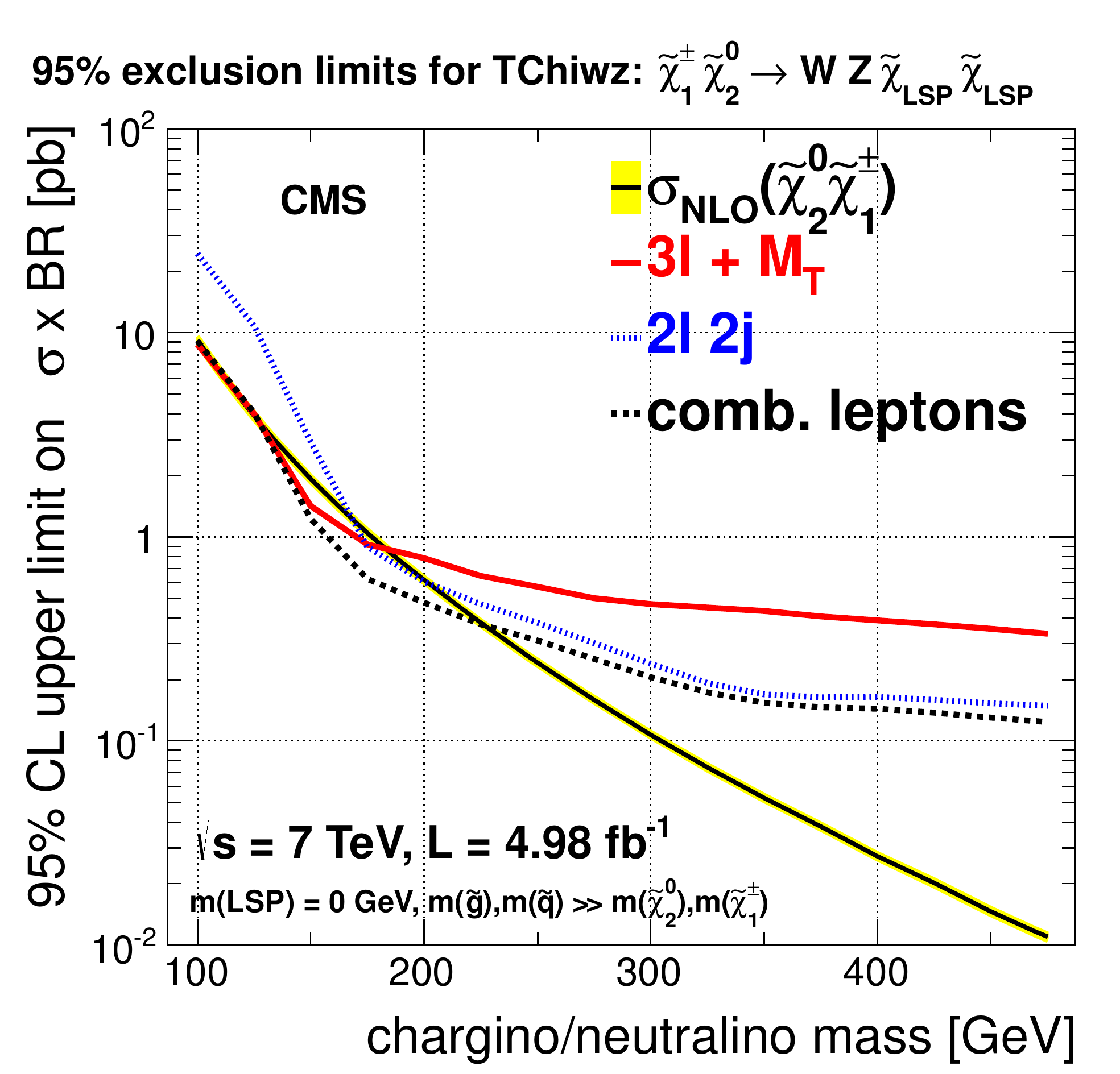}
\caption{The 95\% CL exclusion limits and the predicted cross section for the produced
particle masses with a fixed LSP mass
in the models T1(T1bbbb), T2(T2bb), T1tttt, T2tt, TChiSlepSlep and TChiwz.}
\label{fig:coverage1dPlots}
\end{center}
\end{figure*}

Figure~\ref{fig:coverage1dPlots} shows values of \sigmaXBFUL
for the T1 (T1bbbb), T2 (T2bb), T1tttt,
T2tt, TChiSlepSlep, and TChiwz models
as functions of the produced particle masses at fixed values of the LSP mass.
In the top and middle figures, the LSP mass is fixed at 50\GeV, while in the lower
figures the LSP is fixed to be massless.   Figure~\ref{fig:coverage1dPlots} also illustrates
the method for translating
an upper limit on \sigmaXBFUL to a lower limit on the mass of a hypothetical
particle.   For example, Figure~\ref{fig:coverage1dPlots} (top-left)
displays \sigmaXBFUL for the various analyses that are sensitive to the
T1 and T1bbbb models.   These limits can be compared to \SigNLL for gluino
pair production as a function of gluino mass.
The intersection of \SigNLL with
\sigmaXBFUL determines a lower limit on the gluino mass.
The analyses set a lower limit of approximately $1\TeV$ on the gluino mass
for a LSP mass of $50\GeV$, corresponding
to an upper limit on the cross section of approximately $10\fb$.
This limit assumes ${\cal B}=1$ for the decay of each gluino to
a light-flavor quark-antiquark pair and the LSP.
The (yellow) band on the \SigNLL curve represents an estimate of the
theoretical uncertainties on the cross section calculation.
This figure also demonstrates the decrease in \sigmaXBFUL and the increase on
the upper limit on the gluino mass for those analyses sensitive to the T1bbbb model.

Similar comparisons can be performed for the different
simplified models.
For example, Figure~\ref{fig:coverage1dPlots} (top-right)
displays \sigmaXBFUL for the various analyses that are sensitive to the
T2 and T2bb models.
The analyses set a lower limit of approximately $800\GeV$
on the squark mass
for a LSP mass of $50\GeV$, corresponding
to an upper limit on the cross section of approximately $10\fb$.
This limit assumes there are four squarks with the same mass and
that ${\cal B}=1$ for the decay of each squark to
a light-flavor quark and the LSP.
If only bottom squark-antisquark production is considered, and each bottom squark
decays to a bottom quark and the LSP, a lower limit of approximately $550\GeV$ is
set on the bottom squark mass for a LSP mass of $50\GeV$, corresponding to
an upper limit on the cross section of approximately $20\fb$.
Figure~\ref{fig:coverage1dPlots} (bottom-left) displays the limits on
the model TChiSlepSlep.   A chargino mass of approximately $550\GeV$ is excluded,
corresponding to an upper limit on the cross section of approximately $2\fb$.
This limit assumes that ${\cal B}=1$ for the decay of the chargino and neutralino
to sleptons that further decay to leptons and LSPs.
For the model TChiwz, the limit decreases to $220\GeV$, corresponding to an
upper limit on the cross section of approximately $30\fb$.  This limit
assumes that ${\cal B}=1$ for the decay of the chargino to a \PW~boson and the LSP
and the decay of the neutralino to a \Z~boson and the LSP.

Many of the interpretations
presented in Figure \ref{fig:coveragePlots}
exclude a gluino mass of less than approximately 1\TeV
for a range of LSP masses ranging from 200 to 400\GeV.
However,
the exclusion of a particle mass in a simplified model using SUSY cross sections involves assumptions.
For example, the \SigNLL calculation for gluino pair production
depends upon the choice of squark masses.
If the light-flavor squarks in a specific model, rather than being decoupled, have masses of a few TeV,
the predicted gluino cross sections drop significantly due to destructive interference between
different amplitudes.
The limits on models with cascade decays,
T3w, T5lnu, and T5zz, assume a branching fraction of unity for a gluino decay to a chargino or neutralino.
However, a realistic MSSM model would contain a degenerate chargino-neutralino pair,
reducing the branching fraction to $\frac{1}{2}$ or $\frac{1}{4}$.
Furthermore it should be noted that the lower limits on the sparticle masses have been derived for cross sections
based on the spin assumed in the CMSSM.
Also, the model T2 assumes degenerate copies of left-- and right--handed light-flavor
squarks, while a realistic model may have a significant mass hierarchy between different
squark flavors or eigenstates.
As mentioned earlier, the model T2tt has no spin correlation between the neutralino
and the top quark decay products, while such a correlation will arise in the MSSM
depending on the mixture of interaction quantum states in the mass
quantum states of the top squark and the neutralino.
The information contained in this paper and in the supplementary references
can be used to set limits if any of these assumptions, or others, are
removed or weakened.
It must also be noted that the exclusion limits discussed here only serve to
broadly summarize simplified model results; the full information on the exclusion power of an analysis in the
context of simplified models is contained in the exclusion limits on the production
cross section, as shown in Figure~\ref{fig:T1lhOS_limit}.
This information is contained in the analysis references.
A final caveat can be made regarding the setting of limits in
simplified models.  Since only one signal process is considered,
potential backgrounds are ignored from other signal processes
that may arise in a complete model.

\section{Summary}
\label{sec:conclusions}

The simplified model framework is a recently-developed method for
interpreting the results of searches for new physics.
This paper contains a compilation of simplified model interpretations of CMS supersymmetry analyses
based on 2011 data.  For each simplified model and analysis,
an upper limit on the product of the cross section and
branching fraction is derived as a function of hypothetical particle masses.
Additionally,
lower limits on particle masses are determined
by comparing
the 95\% CL upper limit on the product of the cross section and branching fraction
to the predicted cross section in Supersymmetry for the pair of primary particles.
These lower limits depend upon theoretical assumptions that are described earlier in
this paper.  They should not be regarded as general exclusions on Supersymmetric
particle masses.

The most stringent results for a few simplified models are summarized here.
If the primary particles are gluinos that each decay to quark-antiquark pair and a neutralino,
a gluino of mass of approximately $1\TeV$ is excluded for a neutralino of mass $50\GeV$.
These masses correspond
to an upper limit on the gluino pair production cross section of approximately $10\fb$.
The excluded mass increases if each gluino decays to a bottom quark-antiquark pair and a neutralino, while
the excluded mass decreases if each gluino decays to a top quark-antiquark pair and a neutralino.
The excluded mass also decreases if the gluino undergoes a cascade of decays.
If the primary particles are four squark-antisquark pairs, and each squark
decays to a light-flavor quark and a neutralino, a squark mass of approximately $800\GeV$ is excluded for
a neutralino of mass $50\GeV$, corresponding to an upper limit on the squark-antisquark production
cross section of approximately $10\fb$.
The excluded mass for a single bottom-antibottom squark pair is $550\GeV$.
The comparable exclusion in mass for a single top-antitop squark pair is approximately $150\GeV$ lower.
In the case of the electroweak production of a chargino-neutralino pair, the upper limit
on the cross section is approximately one order of magnitude higher than the corresponding
limit for gluino pair production at the same mass.

The predictions for experimental acceptance and exclusion limits on cross sections
presented here for a range of simplified models and mass parameters
can be used to constrain other theoretical models and compare different analyses.

\section*{Acknowledgments}
\hyphenation{Bundes-ministerium Forschungs-gemeinschaft Forschungs-zentren} We congratulate our colleagues in the CERN accelerator departments for the excellent performance of the LHC and thank the technical and administrative staffs at CERN and at other CMS institutes for their contributions to the success of the CMS effort. In addition, we gratefully acknowledge the computing centres and personnel of the Worldwide LHC Computing Grid for delivering so effectively the computing infrastructure essential to our analyses. Finally, we acknowledge the enduring support for the construction and operation of the LHC and the CMS detector provided by the following funding agencies: the Austrian Federal Ministry of Science and Research; the Belgian Fonds de la Recherche Scientifique, and Fonds voor Wetenschappelijk Onderzoek; the Brazilian Funding Agencies (CNPq, CAPES, FAPERJ, and FAPESP); the Bulgarian Ministry of Education, Youth and Science; CERN; the Chinese Academy of Sciences, Ministry of Science and Technology, and National Natural Science Foundation of China; the Colombian Funding Agency (COLCIENCIAS); the Croatian Ministry of Science, Education and Sport; the Research Promotion Foundation, Cyprus; the Ministry of Education and Research, Recurrent financing contract SF0690030s09 and European Regional Development Fund, Estonia; the Academy of Finland, Finnish Ministry of Education and Culture, and Helsinki Institute of Physics; the Institut National de Physique Nucl\'eaire et de Physique des Particules~/~CNRS, and Commissariat \`a l'\'Energie Atomique et aux \'Energies Alternatives~/~CEA, France; the Bundesministerium f\"ur Bildung und Forschung, Deutsche Forschungsgemeinschaft, and Helmholtz-Gemeinschaft Deutscher Forschungszentren, Germany; the General Secretariat for Research and Technology, Greece; the National Scientific Research Foundation, and National Office for Research and Technology, Hungary; the Department of Atomic Energy and the Department of Science and Technology, India; the Institute for Studies in Theoretical Physics and Mathematics, Iran; the Science Foundation, Ireland; the Istituto Nazionale di Fisica Nucleare, Italy; the Korean Ministry of Education, Science and Technology and the World Class University program of NRF, Republic of Korea; the Lithuanian Academy of Sciences; the Mexican Funding Agencies (CINVESTAV, CONACYT, SEP, and UASLP-FAI); the Ministry of Science and Innovation, New Zealand; the Pakistan Atomic Energy Commission; the Ministry of Science and Higher Education and the National Science Centre, Poland; the Funda\c{c}\~ao para a Ci\^encia e a Tecnologia, Portugal; JINR (Armenia, Belarus, Georgia, Ukraine, Uzbekistan); the Ministry of Education and Science of the Russian Federation, the Federal Agency of Atomic Energy of the Russian Federation, Russian Academy of Sciences, and the Russian Foundation for Basic Research; the Ministry of Science and Technological Development of Serbia; the Secretar\'{\i}a de Estado de Investigaci\'on, Desarrollo e Innovaci\'on and Programa Consolider-Ingenio 2010, Spain; the Swiss Funding Agencies (ETH Board, ETH Zurich, PSI, SNF, UniZH, Canton Zurich, and SER); the National Science Council, Taipei; the Thailand Center of Excellence in Physics, the Institute for the Promotion of Teaching Science and Technology of Thailand and the National Science and Technology Development Agency of Thailand; the Scientific and Technical Research Council of Turkey, and Turkish Atomic Energy Authority; the Science and Technology Facilities Council, UK; the US Department of Energy, and the US National Science Foundation.
Individuals have received support from the Marie-Curie programme and the European Research Council (European Union); the Leventis Foundation; the A. P. Sloan Foundation; the Alexander von Humboldt Foundation; the Belgian Federal Science Policy Office; the Fonds pour la Formation \`a la Recherche dans l'Industrie et dans l'Agriculture (FRIA-Belgium); the Agentschap voor Innovatie door Wetenschap en Technologie (IWT-Belgium); the Ministry of Education, Youth and Sports (MEYS) of Czech Republic; the Council of Science and Industrial Research, India; the Compagnia di San Paolo (Torino); and the HOMING PLUS programme of Foundation for Polish Science, cofinanced from European Union, Regional Development Fund.

\bibliography{auto_generated}
\cleardoublepage \appendix\section{The CMS Collaboration \label{app:collab}}\begin{sloppypar}\hyphenpenalty=5000\widowpenalty=500\clubpenalty=5000\input{SUS-11-016-authorlist.tex}\end{sloppypar}
\end{document}

%% file: SUS-11-016-authorlist.tex
\textbf{Yerevan Physics Institute,  Yerevan,  Armenia}\\*[0pt]
S.~Chatrchyan, V.~Khachatryan, A.M.~Sirunyan, A.~Tumasyan
\vskip\cmsinstskip
\textbf{Institut f\"{u}r Hochenergiephysik der OeAW,  Wien,  Austria}\\*[0pt]
W.~Adam, E.~Aguilo, T.~Bergauer, M.~Dragicevic, J.~Er\"{o}, C.~Fabjan\cmsAuthorMark{1}, M.~Friedl, R.~Fr\"{u}hwirth\cmsAuthorMark{1}, V.M.~Ghete, N.~H\"{o}rmann, J.~Hrubec, M.~Jeitler\cmsAuthorMark{1}, W.~Kiesenhofer, V.~Kn\"{u}nz, M.~Krammer\cmsAuthorMark{1}, I.~Kr\"{a}tschmer, D.~Liko, I.~Mikulec, M.~Pernicka$^{\textrm{\dag}}$, D.~Rabady\cmsAuthorMark{2}, B.~Rahbaran, C.~Rohringer, H.~Rohringer, R.~Sch\"{o}fbeck, J.~Strauss, A.~Taurok, W.~Waltenberger, C.-E.~Wulz\cmsAuthorMark{1}
% ...
\vskip\cmsinstskip
\dag:~Deceased\\
1:~~Also at Vienna University of Technology, Vienna, Austria\\
2:~~Also at CERN, European Organization for Nuclear Research, Geneva, Switzerland\\
% ...
